\documentclass{article}

\usepackage{arxiv}

\title{C++ design patterns for low-latency applications including high-frequency trading}

\author{ \href{https://orcid.org/0000-0001-6846-6649}{\includegraphics[scale=0.06]{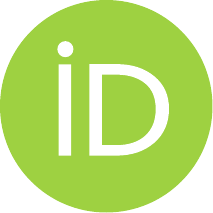}\hspace{1mm}Paul Bilokon} \\
	Departments of Computing and Mathematics\\
	Imperial College London\\
    South Kensington Campus\\
    London SW7 2AZ\\
	\texttt{paul.bilokon@imperial.ac.uk}\\
	\And
	\href{https://orcid.org/0000-0000-0000-0000}{\includegraphics[scale=0.06]{orcid.pdf}\hspace{1mm}Burak Gunduz} \\
	Department of Computing\\
	Imperial College London\\
    South Kensington Campus\\
    London SW7 2AZ\\
	\texttt{burak.gunduz22@imperial.ac.uk} \\
}

\renewcommand{\shorttitle}{C++ design patterns for low-latency applications}

\usepackage{listings}
\lstdefinestyle{customc++}{
    backgroundcolor=\color{white},
    commentstyle=\color{teal},
    keywordstyle=\color{blue},
    numberstyle=\tiny\color{gray},
    stringstyle=\color{purple},
    basicstyle=\ttfamily\small,
    breakatwhitespace=false,         
    breaklines=true,                 
    captionpos=b,                    
    keepspaces=true,                 
    numbers=left,                    
    numbersep=5pt,                  
    showspaces=false,                
    showstringspaces=false,
    showtabs=false,                  
    tabsize=2
}

\usepackage{ifxetex}
\usepackage{textpos}
\usepackage[square,sort,comma,numbers]{natbib}
\usepackage{kpfonts}
\usepackage{ifxetex}
\usepackage{stackengine}
\usepackage{tabularx,longtable,multirow,subfigure,caption}%hangcaption
\usepackage{color}
\usepackage[tight,ugly]{units}
\usepackage{url}
\usepackage{float}
\usepackage[english]{babel}
\usepackage{amsmath}
\usepackage{graphicx}
\usepackage[colorinlistoftodos]{todonotes}
\usepackage{dsfont}
\usepackage{epstopdf} % automatically replace .eps with .pdf in graphics
\usepackage{backref}
\usepackage{array}
\usepackage{latexsym}
\usepackage{etoolbox}

\usepackage{booktabs}

\usepackage{enumerate} % for numbering with [a)] format 

\usepackage{xcolor}

\lstdefinestyle{customc++}{
  language=C++,                          % Specify the language
  backgroundcolor=\color{gray!10},       % Set the background color to light gray
  basicstyle=\ttfamily\small,            % Use small size and monospace font
  keywordstyle=\color{blue}\bfseries,    % Keywords in blue and bold
  commentstyle=\color{olive},            % Comments in a muted green color
  morecomment=[l][\color{magenta}]{\#},  % Preprocessor directives in magenta
  numberstyle=\tiny\color{gray},         % Line numbers in tiny size and gray
  numbers=left,                          % Show line numbers at the left
  numbersep=5pt,                         % Distance of line numbers from code
  showspaces=false,                      % Don't mark spaces
  showstringspaces=false,                % Don't mark spaces in string literals
  showtabs=false,                        % Don't mark tab characters
  tabsize=4,                             % Tab width is 4 spaces
  captionpos=t,                          % Position of caption is at the top
  frame=single,                          % Single frame around code
  rulecolor=\color{black},               % Frame color is black
  breaklines=true,                       % Allow breaking long lines
  breakatwhitespace=true,                % Break lines only at white space
  escapeinside={\%*}{*)}                 % Allow LaTeX commands between %* and *)
}

\ifxetex
\usepackage{fontspec}
\else
\usepackage[pdftex,pagebackref,hypertexnames=false]{hyperref} % provide links in pdf
\usepackage[all]{hypcap}
\fi

\usepackage{tcolorbox}

\usepackage{ntheorem}
\theoremstyle{break}

\allowdisplaybreaks

\makeatletter
\newcounter{elimination@steps}
\newcolumntype{R}[1]{>{\raggedleft\arraybackslash$}p{#1}<{$}}
\def\elimination@num@rights{}
\def\elimination@num@variables{}
\def\elimination@col@width{}

\newcommand{\eliminationstep}[2]
{
    \ifnum\value{elimination@steps}>0\leadsto\quad\fi
    \left[
        \ifnum\elimination@num@rights>0
            \begin{array}
            {@{}*{\elimination@num@variables}{R{\elimination@col@width}}
            |@{}*{\elimination@num@rights}{R{\elimination@col@width}}}
        \else
            \begin{array}
            {@{}*{\elimination@num@variables}{R{\elimination@col@width}}}
        \fi
            #1
        \end{array}
    \right]
    & 
    \begin{array}{l}
        #2
    \end{array}
    &%                                    moved second & here
    \addtocounter{elimination@steps}{1}
}
\makeatother

\makeatletter  
\def\colvec#1{\expandafter\colvec@i#1,,,,,,,,,\@nil}
\def\colvec@i#1,#2,#3,#4,#5,#6,#7,#8,#9\@nil{% 
  \ifx$#2$ \begin{bmatrix}#1\end{bmatrix} \else
    \ifx$#3$ \begin{bmatrix}#1\\#2\end{bmatrix} \else
      \ifx$#4$ \begin{bmatrix}#1\\#2\\#3\end{bmatrix}\else
        \ifx$#5$ \begin{bmatrix}#1\\#2\\#3\\#4\end{bmatrix}\else
          \ifx$#6$ \begin{bmatrix}#1\\#2\\#3\\#4\\#5\end{bmatrix}\else
            \ifx$#7$ \begin{bmatrix}#1\\#2\\#3\\#4\\#5\\#6\end{bmatrix}\else
              \ifx$#8$ \begin{bmatrix}#1\\#2\\#3\\#4\\#5\\#6\\#7\end{bmatrix}\else
                 \PackageError{Column Vector}{The vector you tried to write is too big, use bmatrix instead}{Try using the bmatrix environment}
              \fi
            \fi
          \fi
        \fi
      \fi
    \fi
  \fi 
}  
\makeatother

\robustify{\colvec}

\begin{document}
\maketitle

%%%%%%%%%%%%%%%%%%%%%%%%%%%% Main document

\begin{center}
\vspace*{\fill} % Vertical centering

\begin{abstract}
\noindent This work aims to bridge the existing knowledge gap in the optimisation of latency-critical code, specifically focusing on high-frequency trading (HFT) systems. The research culminates in three main contributions: the creation of a Low-Latency Programming Repository, the optimisation of a market-neutral statistical arbitrage pairs trading strategy, and the implementation of the Disruptor pattern in C++. The repository serves as a practical guide and is enriched with rigorous statistical benchmarking, while the trading strategy optimisation led to substantial improvements in speed and profitability. The Disruptor pattern showcased significant performance enhancement over traditional queuing methods. Evaluation metrics include speed, cache utilisation, and statistical significance, among others. Techniques like Cache Warming and Constexpr showed the most significant gains in latency reduction. Future directions involve expanding the repository, testing the optimised trading algorithm in a live trading environment, and integrating the Disruptor pattern with the trading algorithm for comprehensive system benchmarking. The work is oriented towards academics and industry practitioners seeking to improve performance in latency-sensitive applications. The repository, trading strategy, and the Disruptor library can be found at \href{https://github.com/0burak/imperial_hft}{\textcolor{black}{\texttt{https://github.com/0burak/imperial\_hft}}}.

\end{abstract}

\vspace*{\fill} % Vertical centering
\end{center}

\section{Introduction}
\subsection{Aims}
The overarching aim of this work is to enable individuals to optimise latency-critical code to improve speed. Specifically, the research focuses on programming strategies and data structures relevant to high-frequency trading. Given that the financial industry—particularly buy-side firms focusing on public markets—is secretive, there are not many resources available to guide novice programmers in the field of low-latency programming with a focus on trading. The strategy employed in this research to achieve this goal was to create a bespoke Low-Latency Programming Repository containing various techniques. This was accomplished through extensive research and statistical benchmarking. Using this repository, a systematic trading backtest strategy was optimised. Finally, to provide a holistic view of optimisation techniques for HFT, the Disruptor pattern was implemented in C++ to demonstrate how such data structure can be used in an Order Management Systems (OMS).

\subsection{Research context}
The existing body of literature on optimising various aspects of High-Frequency Trading (HFT) systems primarily originates from industry practitioners who are bound by confidentiality and competitive advantage considerations. Consequently, much of the cutting-edge research in the domain, particularly in areas such as latency improvements, code efficiency, and cache optimisation, remains a closely guarded secret. This underscores the challenge and also the need for academic and open-source initiatives aimed at understanding and improving the inner workings of HFT systems. 

Several works have explored the broader domain of HFT from economic and financial perspectives, attempting to bridge the gap between financial research and computational research. Aldridge focussed on the overarching landscape, system strategies, and economic impact, while publications by Cartea et al. delve into the mathematical models that form the backbone of algorithmic trading systems \cite{2, 7}. However, these works and other literature seldom provide detailed technical insight into aspects like code optimisation or latency reduction. Public talks and conference papers sometimes shed light on HFT systems and their design \cite{3}. Yet, these platforms usually provide a macroscopic view, without direct benchmarking or applying the strategies to relevant systematic trading strategies. When it comes to language-specific research, literature on C++ is comparatively more abundant. However, this literature typically focuses on the design rationale behind C++ and its Standard Template Library, without translating these principles into the specific context of ultra-low-latency HFT systems \cite{8, 9, 10, 11}. In the online domain, blogs and posts sometimes engage with this topic, albeit at a surface level. They often provide averaged latency data, without much analysis into the intricate behaviors affecting cache-access or instruction-execution latencies \cite{17,18,19,20,21}. 

In light of these gaps, the present research aims to make a significant contribution by focusing on the computational and technical aspects of latency optimisation in HFT. This work will not only investigate but also implement and rigorously test various programming strategies within an HFT context, offering insights for both academics and industry practitioners.

\subsection{Summary}
Section 2 explores existing literature and software relevant to the field. Section 2.1 delves into High-Frequency Trading, elucidating its key elements and components such as Data Feed, OMS, Algorithmic Trading Strategies, Risk Management, and Execution Infrastructure. This section also highlights the role of hardware, like FPGAs, in achieving low-latency targets. Section 2.2 evaluates the suitability of the C++ language for HFT, contrasting it with other languages like Java and Rust, and elaborating on specific C++ features beneficial for low-latency applications. Section 2.3 lists and briefly describes various programming strategies and design patterns aimed at optimising C++ code for lower latencies in HFT systems, including Cache Warming, Compile-time Dispatch, and SIMD, among others. Section 2.4 introduces the LMAX Disruptor, a high-performance, low-latency messaging framework designed to minimize contention and latency in concurrent systems. It offers advantages such as efficient memory allocation and optimised sequencing mechanisms for enhanced scalability. Section 2.5 discusses the significance of benchmarking in High-Frequency Trading, specifically addressing the features and limitations of the Google Benchmark library for code performance optimisation in such environments. Section 2.6 outlines the critical role of networking in HFT, detailing technologies and strategies—like fiber-optic communications, colocation, and specialized hardware—geared toward minimizing latency. This section also touches upon the regulatory landscape that shapes these choices. Section 3 contains the Low-Latency Programming Repository. The section is broken down into five sections: compile-time features, optimisation techniques, data handling, concurrency, and system programming. Each section elaborates on their respective programming techniques. Section 4 outlines a market-neutral trading strategy, statistical arbitrage pairs trading, that leverages the mean-reverting nature of price spreads between cointegrated assets. This section focuses on its quantitative foundations, execution methodology, and risk management aspects. It also delves into the critical concept of cointegration, presents the algorithm through a case study involving Goldman Sachs and Morgan Stanley, and discusses CPU-level optimisations for reducing latency. Section 5 details the development and testing of a C++ implementation of the LMAX Disruptor, a high-performance, lock-free inter-thread communication library. This library significantly outperforms traditional queuing methods in both latency and speed by leveraging a ring buffer, sequence numbers, and a specialized waiting strategy. Section 6 evaluates the performance of the Low-Latency Programming Repository, the trading algorithm, and the Disruptor pattern through multiple metrics. These include user feedback, speed improvements, and statistical significance. The section highlights areas for improvement and discusses the impact of latency reduction on trading profitability and risk. Section 7 concludes this work and highlights potential future directions.

\subsection{Contributions}

This work offers both academic and practical contributions to the fields of low-latency programming and high-frequency trading (HFT). The research furnishes comprehensive insights into low-latency programming techniques, as well as the strategies prevalent in the high-frequency trading industry. 

\begin{itemize}
    \item \textbf{Low-Latency Programming Repository:} This repository serves as more than just a theoretical compendium; it is a practical guide enriched by rigorous statistical benchmarking. It offers a curated collection of programming techniques, design patterns, and best practices, all of which are tailored to mitigate latency in HFT systems.

    \item \textbf{Optimisation of Market-Neutral Trading Strategy:} The research led to the successful optimisation of a market-neutral statistical arbitrage pairs trading strategy. This was accomplished through the integration of latency-reduction techniques and CPU-level optimisations. As a result, the trading strategy has shown notable improvements in both execution speed and profitability.

    \item \textbf{Creation of the Disruptor Pattern Library in C++:} This implementation yielded a significant performance enhancement over traditional queuing methods, thus underscoring the real-world applicability and benefits of the Disruptor pattern in latency reduction. It particularly highlights the possible implementation of such a data structure in an HFT system's Order Management System (OMS).
\end{itemize}

\noindent In summary, this work provides contributions in both the research and software library domains, benefiting a broad spectrum of readers ranging from academics to industry practitioners.

\section{Background}
\subsection{HFT}
High-frequency trading (HFT) is an automated trading strategy that utilises technology and algorithms to execute numerous trades at high speeds. Defining HFT precisely can be challenging as it encompasses various aspects of both computer science and finance. The SEC Concept Release on Equity Market Structure outlines five key elements that define this discipline, including high-speed computing, co-location practices, short timeframes for position establishment and liquidation, the submission of multiple promptly followed by cancelled orders, and the objective of concluding the trading day with minimal unhedged positions and near-neutral exposure overnight \cite{1}. HFT systems are mainly built from five main components, including the data feed, order management system (OMS), trading strategies, risk management, and execution infrastructure such as networking. The primary areas of focus in this work will be the trading strategy and OMS. 

The data feed is responsible for receiving and processing real-time market data from various sources, enabling the algorithms to make buy, sell, or  wait decisions. The order management system (OMS) component handles the submission, routing, and monitoring of trade orders, ensuring efficient execution and management of trading activities. Trading strategies form a critical component of HFT systems, as they employ automated algorithms to identify market opportunities, make trading decisions, and execute trades at high speeds. HFT systems encompass four main sections: arbitrage, directional event-based trading, automated market making, and liquidity detection \cite{2}. Within these sections, a variety of trading algorithms are utilised. Statistical arbitrage involves identifying pricing discrepancies between related financial instruments and taking advantage of those discrepancies by simultaneously buying and selling the instruments to profit from the price convergence. Momentum trading strategies focus on exploiting short-term price trends and market momentum, aiming to identify and capitalize on price movements in the direction of the prevailing trend. Pairs trading involves identifying pairs of securities with a historically established correlation and taking simultaneous long and short positions in the two securities, aiming to profit from the convergence or divergence of their prices. Liquidity detection algorithms are designed to identify and trade in illiquid or thinly traded instruments, leveraging advanced data analysis techniques to identify opportunities in less liquid markets or specific trading conditions. News-based Trading algorithms are designed to process and react to news events, such as economic releases or corporate announcements, aiming to exploit market reactions by quickly executing trades based on predefined criteria \cite{3}. Risk management plays a crucial role in HFT systems by implementing measures to assess and mitigate potential risks associated with high-speed trading, ensuring the preservation of capital and minimizing losses. Execution Infrastructure, including networking components, provides the necessary technological framework for low-latency communication and trade execution in HFT systems. Additionally HFT firms make strong use of Field Programmable Gate Arrays (FPGAs) to achieve low-latency targets. FPGAs are a type of programmable circuit which are designed to be configured after manufacturing. FPGAs can execute specific trading algorithms up to 1000 times faster than conventional computers \cite{4}. 
Donadio elucidates that the four key characteristics of FPGAs are programmability, capacity, parallelism, and determinism \citep{Donadio}. First, FPGAs are easily programmable and reprogrammable, capable of handling complex trading algorithms due to their configurable logic blocks (CLBs) and flexible switches. Second, they offer immense capacity with millions of CLBs, although this capacity has physical limitations related to signal propagation times on the silicon die. Third, unlike CPUs that have a fixed architecture and operating system overhead, FPGAs operate on a parallel architecture, allowing multiple code paths to run simultaneously without resource contention, thereby enhancing resilience and speed in trading operations. Lastly, FPGAs offer a high level of determinism, providing repeatable and predictable performance, which is particularly helpful during bursts of market activity. These attributes collectively make FPGAs a powerful tool for HFT, offering advantages in speed, capacity, and reliability. By eliminating software layers and reducing overhead, traders can directly implement algorithms in FPGA hardware, accelerating critical computations and significantly speeding up trading algorithms in HFT. However, this work will primarily focus on software and programming methods for achieving low-latency, excluding discussions on FPGAs and similar hardware solutions.

According to a 2017 estimate by Aldridge and Krawciw, HFT accounted for 10-40\% of trading volume in equities and 10-15\% in foreign exchange and commodities in 2016 \cite{5}. Given the substantial trading volumes and profound impact on financial markets, there is an abundance of information available regarding the economic effects of HFT \cite{6, 7}. However, the literature on the computational and technical aspects of HFT is limited due to the secretive nature of the industry and potential conflicts of interest. Fortunately, there are some insights available from industry experts regarding the requirements for building HFT software, for example (\cite{8, 9, 10, 12, 16, 17, 18}). Carl Cook has covered several C++ optimisation strategies, including Variadic templates, Loop unrolling, Constexpr, and more \cite{10}. This work aims to build upon existing resources and offer the reader a comprehensive compilation of tested data on the performance enhancements achieved through the implementation of these programming strategies within the applied context of HFT.

\subsection{C++}
Low-latency applications such as the ones utilised in HFT tend to prioritise compile-time overhead over runtime overhead, making compiled languages like C++, Java, and Rust more suitable. However, Python is not very well suited for such applications due to its interpreted nature, as it introduces runtime overhead. 
C++ is one of the most popular choices for building software-based HFT systems. This is attributed to the language's fundamental design principles, Ghosh highlights that C++ is preferred for low-latency applications due to its compiled nature, close proximity to hardware, and control over resources, which allow for optimised performance and efficient resource management \cite{Ghosh}. Its rich feature set, including static typing, multiple programming paradigms, and extensive libraries, make it a versatile and highly portable language \cite{14, 15}. However, Rust has emerged as an upcoming alternative to C++ in recent years. Rust offers advantages such as memory safety, concurrency support, developer productivity, performance, and a growing ecosystem, making it an appealing choice for HFT systems development \cite{11}. C++ gained popularity in performance-driven industries due to its ``zero-overhead" principle. This principle is a broad term but it can be witnessed through many ways. For example C++ does not have a garbage collector. This means that programmers must manually allocate and deallocate memory which leads to a more efficient and deterministic memory usage and avoids unnecessary overhead.
However, it's worth noting that Java has been successfully used in latency-critical applications, such as the development of the Disruptor by the LMAX Group, as discussed in greater detail in Section 2.4. Java mitigates the impact of garbage collection in the Disruptor pattern by preallocating and reusing objects within the ring buffer, minimizing frequent memory allocation and subsequent garbage collection pauses. Under specific circumstances, Java can exhibit superior performance compared to C++, particularly when the virtual machine and garbage collector are carefully optimised to align with the application's garbage collection interrupt frequency. Such fine-tuning can lead to Java outperforming C++ in certain scenarios, as observed by Chandler Carruth \cite{12}. Furthermore, Aldrige highlights the example of the Nasdaq OMX system, which is developed in Java with the garbage collector disabled \cite{2}. In contrast to Java, Rust does not employ a garbage collector. Instead, Rust utilises its unique ownership system and ``ownership and borrowing" concept to manage memory. The ownership system ensures that each piece of memory has a single owner at any given time, and Rust automatically frees the memory when the owner goes out of scope. This approach provides a middle ground between the memory management models of C++ and Java.

Additionally, C++ offers various techniques to shift processing tasks from runtime to compile-time. For instance, templates enable developers to move the runtime overhead associated with dynamic polymorphism to compile-time in exchange for flexibility. The ``Curiously Recurring Template Pattern" is a notable example in this regard.
Another technique utilised in C++ is the use of inline functions, which allows the compiler to directly insert the function's code at the call site. This reduces the overhead of a standard function call. Finally, the ``constexpr" keyword can be used to evaluate computations at compile time instead of runtime, resulting in quicker and more efficient code execution. These techniques contribute to C++'s versatility and enable optimisation at the compile-time level, enhancing performance. As C++ is one of the most suitable languages for HFT, it will be the focus of this work.
However, it is important to consider that factors such as the compiler (and its version), machine architecture, 3rd party libraries, build and link flags can also affect latency. Therefore, it is crucial to examine the machine instructions produced by C++. To accomplish this, Compiler Explorer will be utilised. Compiler Explorer, designed by Matt Godbolt, is an online platform that enables users to explore and analyze the output generated by various compilers for different programming languages. It provides an interface for writing code and viewing the corresponding compiled assembly or intermediate representations of the code. Rasovsky discussed the challenges of benchmarking time-sensitive operations in a high-frequency trading environment. Rasovsky initially found that the commonly used clock\_gettime system call appeared to be 85 times slower than using Intel's Time Stamp Counter (TSC) on a benchmarking website, but upon running tests on her actual server, the gap narrowed to only twice as slow \cite{rasovsky}. The talk emphasizes the importance of context-sensitive benchmarking and reveals how different environments and underlying system configurations can dramatically impact performance metrics.

\subsection{Design patterns}
Design patterns and programming strategies will be used interchangeably throughout this work. As mentioned earlier there are many ways to optimise C++ code to achieve lower latencies. The strategies that will be examined in this work are as follows.

\begin{itemize}
    \item \textit{Cache Warming:} To minimize memory access time and boost program responsiveness, data is preloaded into the CPU cache before it's needed \cite{9}.
    \item \textit{Compile-time Dispatch:} Through techniques like template specialization or function overloading, optimised code paths are chosen at compile time based on type or value, avoiding runtime dispatch and early optimisation decisions.
    \item \textit{Constexpr:} Computations marked as constexpr are evaluated at compile time, enabling constant folding and efficient code execution by eliminating runtime calculations \cite{10}.
    \item \textit{Loop Unrolling:} Loop statements are expanded during compilation to reduce loop control overhead and improve performance, especially for small loops with a known iteration count.
    \item \textit{Short-circuiting:} Logical expressions cease evaluation when the final result is determined, reducing unnecessary computations and potentially improving performance.
    \item \textit{Signed vs Unsigned Comparisons:} Ensuring consistent signedness in comparisons avoids conversion-related performance issues and maintains efficient code execution.
    \item \textit{Avoid Mixing Float and Doubles:} Consistent use of float or double types in calculations prevents implicit type conversions, potential loss of precision, and slower execution.
    \item \textit{Branch Prediction/Reduction:} Accurate prediction of conditional branch outcomes allows speculative code execution, reducing branch misprediction penalties and improving performance.
    \item \textit{Slowpath Removal:} Optimisation technique aiming to minimize execution of rarely executed code paths, enhancing overall performance.
    \item \textit{SIMD:} Single Instruction, Multiple Data (SIMD) allows a single instruction to operate on multiple data points simultaneously, significantly accelerating vector and matrix computations.
    \item \textit{Prefetching:} Explicitly loading data into cache before it is needed can help in reducing data fetch delays, particularly in memory-bound applications.
    \item \textit{Lock-free Programming:} Utilises atomic operations to achieve concurrency without the use of locks, thereby eliminating the overhead and potential deadlocks associated with lock-based synchronization.
    \item \textit{Inlining:} Incorporates the body of a function at each point the function is called, reducing function call overhead and enabling further optimisation by the compiler.
\end{itemize}

\subsection{LMAX Disruptor}
The LMAX Disruptor was created by LMAX Exchange to address the specific requirements of their high-performance, low-latency trading system. The LMAX Disruptor offers a highly optimised and efficient messaging framework that enables concurrent communication between producers and consumers with minimal contention and latency \cite{13}. Traditional multithreaded applications often suffer from contention issues when multiple threads try to access shared resources concurrently. Locking mechanisms, such as mutexes or semaphores, are commonly used to synchronize access to shared data structures, but they can introduce significant overhead and contention, leading to poor scalability and increased latency. This is because locks require arbitration when contended, and this arbitration is achieved by a context switch. However, context switches are costly as the cache associated with the previous thread or process may be invalidated, causing cache misses. The cache might also need to be flushed, resulting in additional latency. Finally, when a new thread or process starts, it takes time to populate the cache with relevant data, affecting performance initially \cite{13}.

Another way to tackle contention is the use of Compare And Swap (CAS) operations. CAS is an atomic instruction used in concurrent programming to implement synchronization and ensure data integrity in multi-threaded environments. It is a fundamental building block for lock-free and wait-free algorithms. Firstly, each thread involved in the CAS operation reads the current value of the shared variable from memory. Next, the thread compares this current value with its expected value, which it has in mind. If the current value matches the expected value, indicating that no other thread has modified the shared variable since the read operation, the thread proceeds to atomically swap the current value with a new desired value. Following the swap, the CAS operation returns a status indicating whether the swap was successful or not. If successful, it signifies that the thread has successfully updated the shared variable; otherwise, it means that another thread had modified the shared variable in the meantime, requiring the thread to retry the CAS operation. CAS is definitely better than the use of locks, but they're still not cost-free. Firstly, orchestrating a complex system using CAS operations can be harder than the use of locks. Secondly, to ensure atomicity, the processor locks its instruction pipeline, and a memory barrier is used to ensure that changes made by a thread become visible to other threads \cite{13}. Queues can be implemented using linked lists or arrays, but unbounded queues can lead to memory exhaustion if producers outpace consumers. To prevent this, queues are often bounded or actively tracked in size. However, bounded queues introduce write contention on head, tail, and size variables, leading to high contention and potential cache coherence issues. Managing concurrent access to queues becomes complex, especially when dealing with multiple producers and consumers.

The LMAX Disruptor aims to resolve the mentioned issues and optimise memory allocation efficiency while operating effectively on modern hardware. It achieves this through a pre-allocated bounded data structure called a ring buffer. Producers add data to the ring buffer, and one or more consumers process it. According to LMAX's benchmark results, they state that the Disruptor achieves significantly lower mean latency, by three orders of magnitude, compared to an equivalent queue-based approach in a three-stage pipeline \cite{13}. Additionally, the Disruptor demonstrates approximately 8 times higher throughput handling capacity \cite{13}. The ring buffer can store either an array of pointers to entries or an array of structures representing the entries. Typically, each entry in the ring buffer does not directly contain the data being passed but serves as a container or holder for the data. As the LMAX Disruptor was built on Java, the garbage collector was an issue. As mentioned earlier, the ring buffer is pre-allocated, meaning the entries in the buffer do not get cleaned up by the garbage collector, and they exist for the lifetime of the Disruptor. In most use cases of the Disruptor, there is typically only one producer, such as a file reader or network listener, resulting in no contention on sequence/entry allocation. However, in scenarios with multiple producers, they race to claim the next available entry in the ring buffer, which can be managed using a CAS operation on the sequence number. Once a producer copies the relevant data to the claimed entry, it can make it public to consumers by committing the sequence. Consumers wait for a sequence to become available before reading the entry, and various strategies can be employed for waiting, including condition variables within a lock or looping with or without thread yield. The Disruptor avoids CAS contention present in lock-free multi-producer-multi-consumer queues, making it a scalable solution.

Sequencing is a fundamental concept in the Disruptor for managing concurrency. Producers and consumers interact with the ring buffer based on sequencing. Producers claim the next available slot in the sequence, which can be a simple counter or an atomic counter with CAS operations. Once a slot is claimed, the producer can write to it and update a cursor representing the latest entry available to consumers. Consumers wait for a specific sequence by using memory barriers to read the cursor, ensuring visibility of changes. Consumers maintain their own sequence to track their progress and coordinate work on entries. In cases with a single producer, locks or CAS operations are not needed, and concurrency coordination relies solely on memory barriers. The Disruptor offers an advantage over queues when consumers wait for an advancing cursor in the ring buffer. If a consumer notices that the cursor has advanced multiple steps since it last checked, it can process entries up to that sequence without involving concurrency mechanisms. This allows the consumer to quickly catch up with producers during bursts, balancing the system. This batching approach improves throughput, reduces latency, and provides consistent latency regardless of load until the memory system becomes saturated.

The Disruptor framework can be seen in Figure 1. It simplifies the programming model by allowing producers to claim entries, write changes into them, and commit them for consumption via a ProducerBarrier. Consumers only need to implement a BatchHandler to receive callbacks when new entries are available. This event-based programming model shares similarities with the Actor Model.

\begin{figure}[H]
      \centering
      \includegraphics[width=1\textwidth]{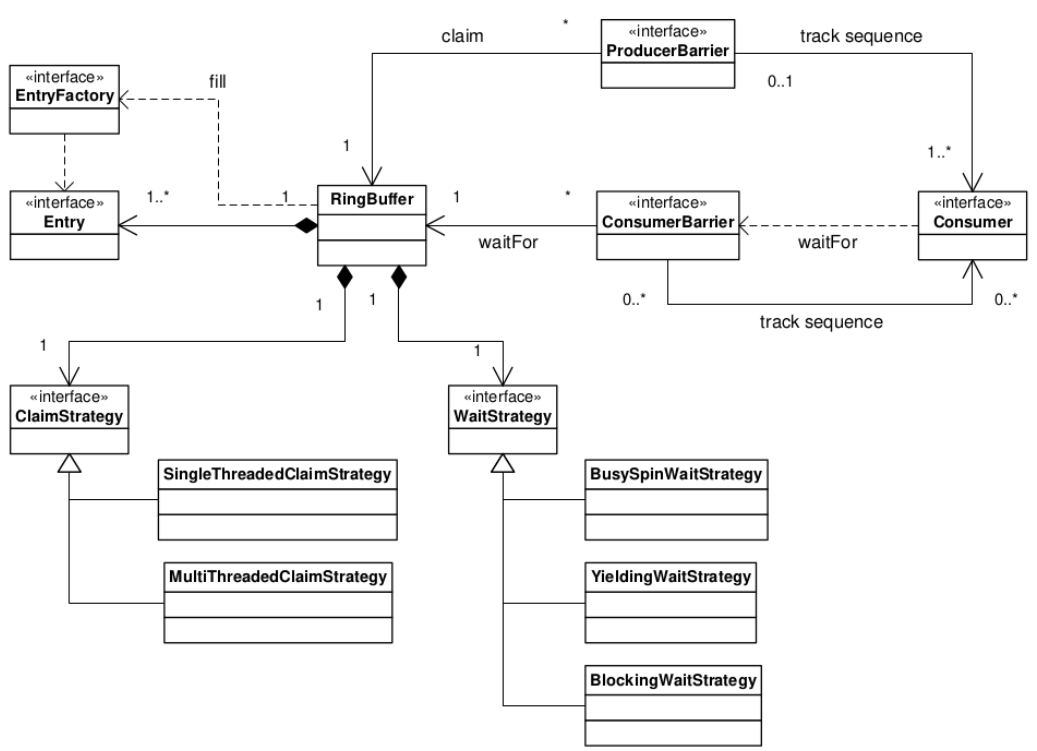}
      \caption{UML of Disruptor Framework \cite{13}.}
      \label{fig:sample-image}
\end{figure}

\noindent By separating concerns that are typically intertwined in queue implementations, the Disruptor enables a more flexible design. The central component is the RingBuffer, which facilitates data exchange without contention. Producers and consumers interact with the RingBuffer independently, with concurrency concerns handled by the ProducerBarrier for slot claiming and dependency tracking, and the ConsumerBarrier for notifying consumers of new entries. Consumers can be organized into a dependency graph, representing multiple stages in a processing pipeline.

\subsection{Benchmarking}
In the realm of HFT, where decisions are made in microseconds, optimising the code performance becomes an absolute necessity. Given the spped at which trades are conducted, even minor improvements can have significant impacts on the overall performance and profitability. The challenge, therefore, lies in effectively identifying, analysing, and implementing these enhancements, a task that necessitates the deployment of robust profiling and benchmarking tools. In this work two main tools were used, Google Benchmark for speed and perf (Linux) for cache analysis.

\subsubsection*{Google Benchmark}
Google Benchmark is an open-source library designed for benchmarking the performance of code snippets in C++ applications. However, it is critical to evaluate the appropriateness of Google Benchmark for HFT environments. One of the standout features of Google Benchmark is its use of statistically rigorous methodology. The library performs multiple iterations of specific code snippets or functions, thereby generating accurate performance metrics. Such precision in performance measurement is critical in the context of HFT, where the software stack's efficiency can directly influence the trading strategy's efficacy. Furthermore, Google Benchmark offers an impressive versatility with its support for multiple benchmarking modes, including CPU-bound and real-time modes. This can, in turn, guide the code optimisation strategies to meet the specific requirements of applications. However, it is important to note that while Google Benchmark excels in providing detailed performance metrics, it might not cover some specifics unique to HFT scenarios, such as network latency, co-location effects, and hardware timestamping, among others. It primarily focuses on CPU-bound tasks, which may not always reflect the entire performance landscape in HFT \cite{Lilja}. Therefore, while Google Benchmark serves as a powerful tool in optimising code and improving computational efficiency, it should be supplemented with other performance analysis tools that can monitor and optimise these HFT-specific factors.
\subsubsection*{perf (Linux)}
The \texttt{perf} tool, also known as Performance Counters for Linux (PCL), is a sophisticated performance monitoring utility that is integrated into the Linux kernel. It provides a rich set of commands and options for profiling and tracing software performance and system events at multiple layers, from hardware-level instruction cycles to application-level function calls. For this work perf was utilised for cache analysis, in particular cache misses.

\subsection{Cache analysis}
In computer science, a cache is a hardware or software component that stores a subset of data, typically transient in nature, so that future requests for that data can be served more rapidly \cite{Smith}. The data most recently or frequently accessed is usually stored in the cache, as subsequent reads or writes to the same data can often be completed more quickly. By storing data closer to the processor or the end-user, the cache reduces latency and improves I/O efficiency, thus accelerating data access and enhancing overall system performance. Caches are implemented in various levels of a computing architecture, including but not limited to, CPU caches (L1, L2, L3), disk caches, and even distributed network caches \cite{Solihin}.
The terms ``cache hit" and ``cache miss" refer to the efficacy of data retrieval from the cache. A cache hit occurs when a data request can be fulfilled by the cache, negating the need to access the slower, underlying data store, whether that be main memory, disk storage, or a remote server \cite{Jouppi}. This results in a faster response time and less resource utilisation for the specific data operation. Conversely, a cache miss signifies that the requested data is not present in the cache, necessitating a fetch from the slower data store to both fulfill the current request and update the cache for potential future accesses \cite{Jouppi}. Cache misses are generally more expensive in terms of time and computational resources, and they can occur for various reasons, including the first-time access of data (cold start), eviction of data due to cache size limitations, or an ineffective cache management strategy \cite{Zhong}. Figure 2 outlines this concept. The ratio of cache hits to total access attempts is often used as an important metric for evaluating the effectiveness of a cache system, and is a metric which is used in this work for evaluation of both the trading strategy and the Disruptor pattern.  

\begin{figure}[H]
      \centering
      \includegraphics[width=0.65\textwidth]{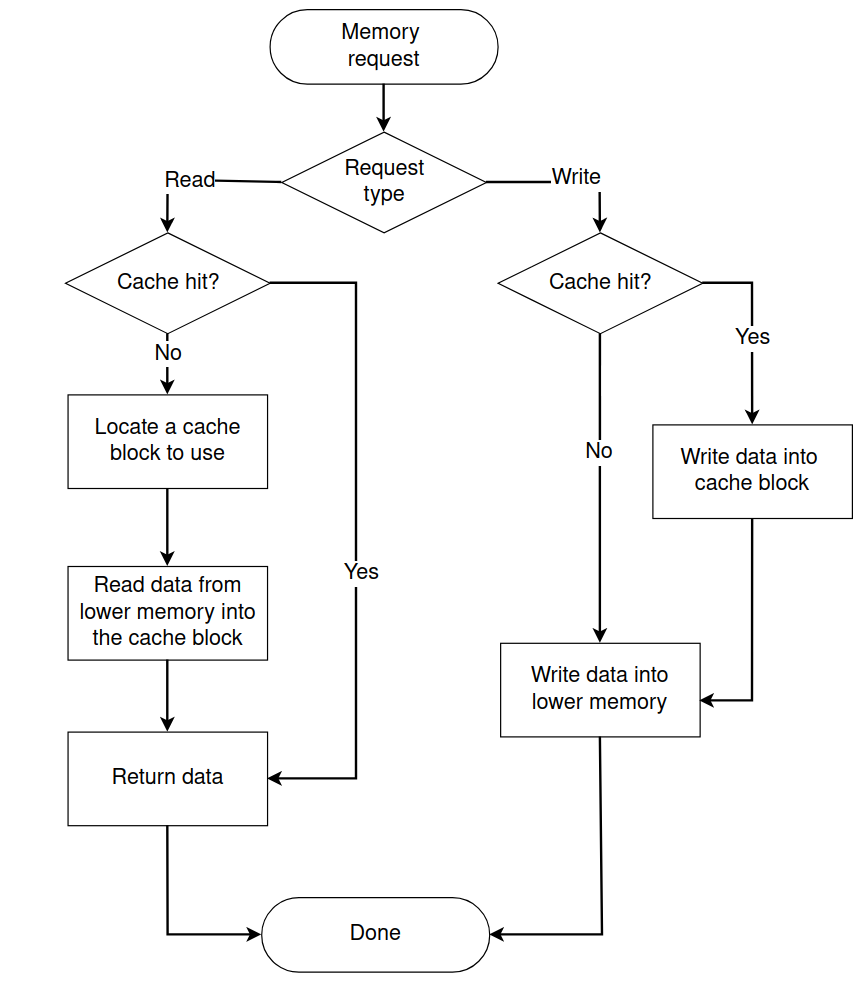}
      \caption{Flowchart of a memory request showing cache hit and miss.}
      \label{fig:sample-image}
\end{figure}

\subsection{Networking}

Low-latency networks are the linchpin of high-frequency trading. Although networking is not a key aspect of this work it is still important to inform the reader with the basic understanding of the surrounding pieces of a complex HFT system. The propagation delays in transmitting information can have immediate financial implications. As such, most HFT firms opt for fiber-optic communication to enable data transmission at speeds close to the speed of light \cite{2}. However, firms are continuously researching ways to shave off additional microseconds. One such method is through the use of microwave and millimeter-wave communications. Such technologies bypass the physical limitations of fiber-optic cables and can transmit data even faster over shorter distances \cite{19}. Colocation remains a crucial strategy for HFT firms, providing them the benefit of proximity to a stock exchange’s data center, thus further minimizing latency \cite{20,21}. The advantages gained through colocation have led to its commercialization, with exchanges now charging significant fees for premium colocation services \cite{22}. Beyond networking protocols and physical infrastructure, HFT firms invest in specialized hardware. Field-Programmable Gate Arrays (FPGAs) and Application-Specific Integrated Circuits (ASICs) are common choices, as they can execute trading algorithms more efficiently than general-purpose processors \cite{23, 24}. Another critical aspect is the network topology used within the trading infrastructure. The design of these networks focuses on maximizing speed and minimizing the number of hops between network nodes. This often involves direct connections between key components in the trading infrastructure, thereby avoiding potential points of latency and failure \cite{2, 25}. While speed is of the essence, the robustness and security of the network cannot be compromised. Redundancy measures are usually put in place, including dual network paths, failover systems, and backup data centers to maintain a 100\% uptime \cite{26, 2}. The rising threats of cyber-attacks also necessitate that HFT firms employ robust security protocols to protect their systems \cite{27}. Continuous monitoring and real-time analytics are also crucial for maintaining optimal performance. HFT firms utilise network monitoring solutions that can detect and alert in real-time if the latency exceeds predetermined thresholds \cite{28,29}. Additionally time is a critical factor in high-frequency trading, and the use of precise time-synchronization protocols like Precision Time Protocol (PTP) is becoming increasingly important. Accurate timestamping of events allows for fairer market conditions and is often a regulatory requirement \cite{32}. Advancements in networking technology have attracted the attention of regulators concerned about market fairness. Regulations like the U.S. SEC's Regulation NMS (National Market System) aim to foster both innovation and competition while maintaining a level playing field for investors \cite{30, 31}. In summary, the networking technology in HFT is an intricate web of hardware, software, and topological design optimised for speed and reliability. It is a dynamic field with continuous innovation, driven by the relentless pursuit of a competitive advantage and also shaped by regulatory considerations.

\section{Low-Latency Programming Repository}

This section contains the Low-Latency Programming Repository, whichc can be accessed from \href{https://github.com/0burak/imperial_hft}{\textcolor{black}{\texttt{https://github.com/0burak/imperial\_hft}}}. The programming techniques are aggregated into five main categories: compile-time features, optimisation techniques, data handling, concurrency, and system programming.

\subsection{Compile-Time Features}
\subsubsection*{Cache Warming}
Cache warming, in computing, refers to the process of pre-loading data into a cache memory from its original storage disk, with the intent to accelerate data retrieval times. The logic behind this practice is that accessing data directly from the cache, which is considerably faster than the hard disk or even solid-state drives, significantly reduces latency and improves overall system performance. By pre-loading or ``warming up" the cache, the system is prepared to serve the required data promptly when it is needed, bypassing the need to retrieve it from the slower primary storage. Figure 3 highlights the speed difference for accessing data in different levels of the cache. This technique is particularly useful when the data access pattern is predictable, allowing specific data to be pre-loaded efficiently into the cache. 

\begin{figure}[H]
      \centering
      \includegraphics[width=1\textwidth]{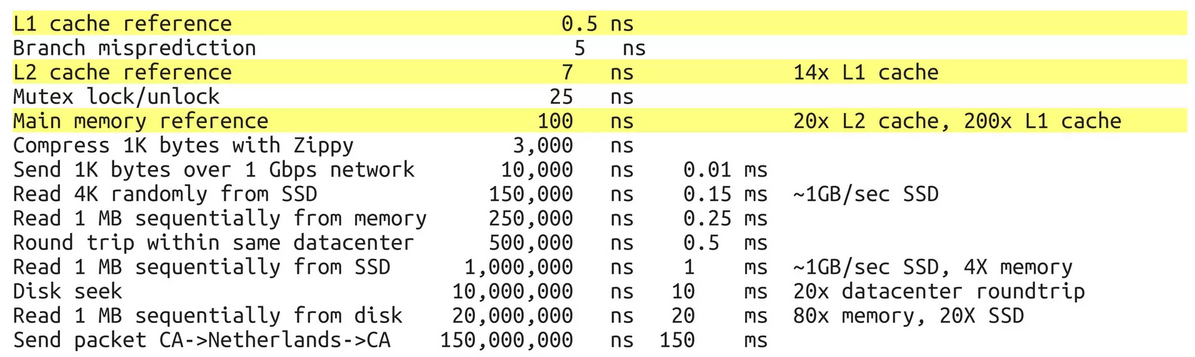}
      \caption{Speed  comparison of fetching data from L1, L2, and main memory \cite{33}.}
      \label{fig:sample-image}
\end{figure}

\noindent In HFT, the execution path that reacts to a trade signal is known as the hot path \cite{2}. Although the hot path isn't triggered frequently, when it is, it's vital for it to run swiftly due to the extremely tight time frames HFT operates on. To achieve this speed, cache warming is utilised. Cache warming pre-loads the necessary data and instructions for the hot path into the cache \cite{Weller, 9}. This means when a trade signal is detected, the system can immediately access the required data from the fast cache memory, rather than slower storage mediums, significantly reducing latency and potentially increasing the success rate of the trade. This approach leverages the waiting periods in HFT to prepare, or warm, the cache for instantaneous action once a trading signal occurs. As observed in Figure 4, the execution engine code is not executed every time therefore when it needs to be executed the cache is not warm. However in Figure 5, the execution engine code is executed every single time but only released the order when a trade must be placed. This keeps the cache ``warm".

\begin{figure}[H]
      \centering
      \includegraphics[width=0.8\textwidth]{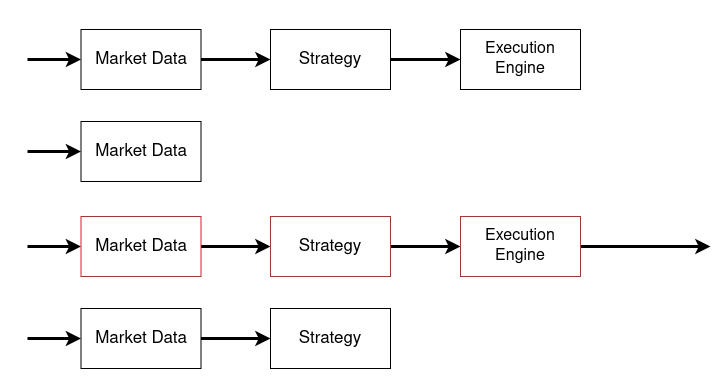}
      \caption{Flowchart of ``cold'' cache.}
      \label{fig:sample-image}
\end{figure}

\begin{figure}[H]
      \centering
      \includegraphics[width=0.8\textwidth]{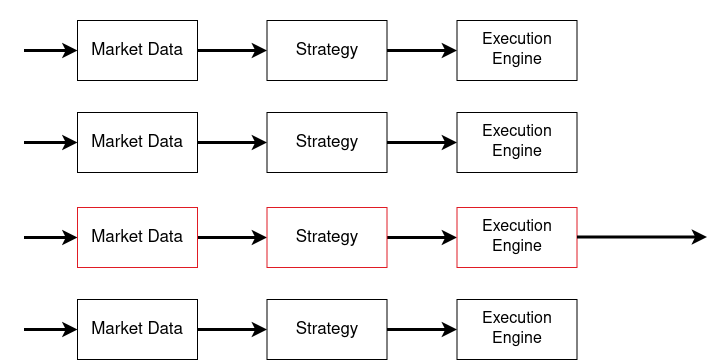}
      \caption{Flowchart of ``warm" cache..}
      \label{fig:sample-image}
\end{figure}

 \noindent Two scenarios are tested: \texttt{BM\_CacheCold} and \texttt{BM\_CacheWarm}. In \texttt{BM\_CacheCold}, data is accessed in a random order, creating a situation where the cache is ``cold''---data has to be fetched from the main memory, which incurs significant latency. In \texttt{BM\_CacheWarm}, the cache is ``warmed up'' by accessing data in a sequential order. In this scenario, the data is already present in the cache (due to the spatial locality principle), resulting in faster access times. The \texttt{benchmark::DoNotOptimise} function is used to prevent the compiler from optimizing the code during the warming-up and testing process.

 \begin{lstlisting}[style=customc++, language=C++, caption={Cold cache Google benchmark}]
  static void BM_CacheCold(benchmark::State& state) {
  // Generate random indices
  for(auto& index : indices) {
    index = rand() % kSize;
  }
  for (auto _ : state) {
    int sum = 0;
    // Access data in random order
    for (int i = 0; i < kSize; ++i) {
      benchmark::DoNotOptimise(sum += data[indices[i]]);
    }
    benchmark::ClobberMemory();
  }
}   
\end{lstlisting}

\begin{lstlisting}[style=customc++, language=C++, caption={Warm cache Google benchmark}]
static void BM_CacheWarm(benchmark::State& state) {
  // Warm cache by accessing data in sequential order
  int sum_warm = 0;
  for (int i = 0; i < kSize; ++i) {
    benchmark::DoNotOptimise(sum_warm += data[i]);
  }
  benchmark::ClobberMemory();
 
  // Run the benchmark
  for (auto _ : state) {
    int sum = 0;
    // Access data in sequential order again
    for (int i = 0; i < kSize; ++i) {
      benchmark::DoNotOptimise(sum += data[i]);
    }
    benchmark::ClobberMemory();
  }
}
\end{lstlisting}

\noindent From the benchmark results, it can be seen that the warm cache scenario (\texttt{BM\_Cache\-Warm}) performed significantly better than the cold cache scenario (\texttt{BM\_CacheCold}). The cold cache test took approximately 267,685,006 nanoseconds, while the warm cache test took about 25,635,035 nanoseconds. This indicates that cache warming resulted in an immense speed improvement of about \(90\%\). This dramatic improvement is due to the fact that, with cache warming, the data required for computations is readily available in the cache, eliminating the need to fetch it from the main memory. Therefore, these results underscore the value and efficiency of cache warming in scenarios where data access patterns are predictable and regular.

To further examine the results, \texttt{perf} was used to analyse cache utilisation behaviour. The results can be seen in Table 1. The \texttt{Cache Cold} scenario entailed accessing elements from a large array in a random order, thereby demonstrating poor spatial and temporal locality. This resulted in a high number of cache misses, with approximately \(73.96\%\) of all cache references leading to a miss. On the other hand, the \texttt{BM\_CacheWarm} scenario involved a pre-pass that accessed data in a sequential manner before the timed loop. This access pattern leveraged spatial locality, effectively reducing the total number of cache references. Interestingly, the cache miss rate remained relatively stable in both scenarios---\(73.96\%\) in \texttt{BM\_CacheCold} versus \(71.55\%\) in \texttt{BM\_CacheWarm}. However, the total number of cache references was lower in the \texttt{BM\_CacheWarm} case, thereby illustrating more efficient cache utilisation.

The difference in execution time between the two scenarios further emphasizes the impact of cache behavior. The \texttt{BM\_CacheWarm} scenario exhibited significantly lower execution time due to its more efficient use of cache, despite the large data set that likely exceeded any cache level's capacity. The instruction count also increased substantially in the \texttt{BM\_CacheWarm} case, which can be attributed to the higher number of iterations executed in the same loop owing to the time saved by reduced cache misses. This experiment underscores the critical role that cache access patterns play in affecting program performance, even when the cache miss rate exhibits only minor variations. 

\begin{table}[htbp]
\centering
\begin{tabular}{lrrr}
\toprule
\textbf{Metrics} & \textbf{BM\_CacheCold} & \textbf{BM\_CacheWarm} \\
\midrule
Speed (ns)     & 267,685,006  & 25,635,035  \\
Instructions     & 4,931,929,489 & 12,013,354,366   \\
Cache References   & 146,264,562  & 61,306,992   \\
Cache Misses (\% of all cache refs)& 73.964 & 71.559  \\
\bottomrule
\end{tabular}
\caption{Comparison between \texttt{BM\_CacheCold} and \texttt{BM\_CacheWarm}}
\label{tab:comparison}
\end{table}

\subsubsection*{Compile-time dispatch}
Runtime dispatch and compile-time dispatch are two techniques in object-oriented programming that determine which specific function gets executed \cite{grothoff}. Runtime dispatch, also known as dynamic dispatch, resolves function calls at runtime. This method is primarily associated with inheritance and virtual functions \cite{cardelli}. In such cases, the function that gets executed relies on the object's type at runtime. Conversely, \texttt{compile-time dispatch} determines the function call during the compilation phase and is frequently used in conjunction with templates and function overloading. Since the dispatch decision occurs during compilation, it typically results in swifter execution, eliminating the runtime overhead.

In the benchmark results, a distinct performance difference emerges between runtime and compile-time dispatch. For the runtime dispatch test, \texttt{BM\_RuntimeDis\-patch}, times were \(2.60\, \text{ns}\) and \(2.15\, \text{ns}\) for \texttt{Derived1} and \texttt{Derived2} respectively. In contrast, the compile-time dispatch test, \texttt{BM\_CompileTimeDispatch}, recorded a more efficient time of \(1.92\, \text{ns}\) for both \texttt{Derived1} and \texttt{Derived2}. These data points underscore the efficiency of compile-time dispatch, which shaves off approximately \(0.68\, \text{ns}\) and \(0.23\, \text{ns}\) in execution time respectively. The noted efficiency in compile-time dispatch is due to decisions about function calls being made during the compilation phase. By bypassing the decision-making overhead present in runtime dispatch, programs can execute more swiftly, thus boosting performance.

\subsubsection*{Constexpr}
Constexpr is a keyword in C++ facilitating the evaluation of expressions during compilation rather than runtime \cite{14}. Its advantage lies in enhancing performance by computing values at compile time, thus obviating the necessity for runtime computations. This is particularly helpful for constant expressions with values determinable at compile time. In the present study, an experiment delved into the performance contrast between two computation methodologies: computation at compile-time using C++'s constexpr keyword and conventional runtime computation. Google Benchmark tool gauged the time necessary to ascertain the factorial of 10 through two analogous functions: one utilising constexpr and another adhering to standard runtime computations.

\begin{lstlisting}[style=customc++, language=C++, caption={Constexpr Google Benchmark}]
constexpr int factorial(int n) {
	return (n <= 1) ? 1 : (n * factorial(n - 1));
}

int runtime_factorial(int n) {
	return (n <= 1) ? 1 : (n * runtime_factorial(n - 1));
}

static void BM_ConstexprFactorial(benchmark::State& state) {
	for (auto _ : state) {
    	benchmark::DoNotOptimise(factorial(10));
	}
}
BENCHMARK(BM_ConstexprFactorial);

static void BM_RuntimeFactorial(benchmark::State& state) {
	for (auto _ : state) {
    	benchmark::DoNotOptimise(runtime_factorial(10));
	}
}
BENCHMARK(BM_RuntimeFactorial);

BENCHMARK_MAIN();

\end{lstlisting}

\noindent Both the constexpr function, termed factorial, and its parallel, \texttt{runtime\_factorial}, are functionally congruent, employing recursion for factorial computations. Their distinctive feature is the incorporation of the constexpr keyword in the former, informing the compiler of its potential for compile-time computations. Performance evaluation hinged on two benchmark functions, \texttt{BM\_ConstexprFactorial} and \texttt{BM\_Ru\-ntimeFactorial}. These functions invoke the factorial and \texttt{runtime\_factorial} functions within a repetitive framework. The \texttt{benchmark::DoNotOptimise} function was integrated within both benchmarks to deter the compiler from early optimisation of the factorial calculation, ensuring accurate performance metrics. Data showcased a notable performance gap between the factorial functions. The constexpr function recorded a computational span of approximately \(0.245\) nanoseconds per iteration, while its runtime counterpart necessitated a heftier \(2.69\) nanoseconds per iteration. These results suggest the constexpr function's speed exceeded its runtime counterpart by about \(90.88\%\) in computing the factorial of 10. However, it's imperative to understand that this marked performance variance doesn't imply constexpr consistently amplifies runtime speed. The primary objective of the constexpr keyword is to shift computations from runtime to compile-time, not necessarily to boost runtime velocity.

It's also pivotal to acknowledge that compilers aren't inhibited from optimising either constexpr or non-constexpr code. Consequently, in various instances, discernible runtime performance disparities might remain elusive. The pronounced performance elevation observed in the constexpr function in this study can be ascribed predominantly to the distinct way the compiler addressed constexpr and optimisation nuances.

\subsubsection*{Inlining}
In computer programming, inlining refers to a compiler optimisation method in which a function call is replaced by the actual content of the function \cite{14}. This procedure aims to reduce the overhead typically linked with function calls, such as parameter transmission, stack frame handling, and the function call-return process. The directive \texttt{\_\_attribute\_\_((always\_inline))} is a particular instruction for the GNU Compiler Collection (GCC), guiding the compiler to persistently inline the designated function, irrespective of the optimisation level chosen for the compilation.

The benchmark test in the given code contrasts two analogous functions: \texttt{add} and \texttt{add\_inline}. The latter bears the \texttt{\_\_attribute\_\_((always\_inline))} directive. Results indicate that the function utilising the inlining directive (\texttt{BM\_WithInline}) exhibited faster performance, averaging \(1.90\) nanoseconds per operation. In contrast, the function without inlining (\texttt{BM\_WithoutInline}) registered \(2.39\) nanoseconds per operation. This denotes a speed enhancement of approximately \(20.5\%\). The performance enhancement can be attributed to the \texttt{always\_inline} function, which sidesteps the overhead linked to function calls, thereby curtailing latency. These findings underscore the potential performance advantages of inlining, particularly for compact functions that are invoked frequently. However, it's imperative to approach this method with discretion, as over-reliance on inlining may inflate the binary size, possibly undermining instruction cache efficacy.

\begin{figure}[H]
      \centering
      \includegraphics[width=0.9\textwidth]{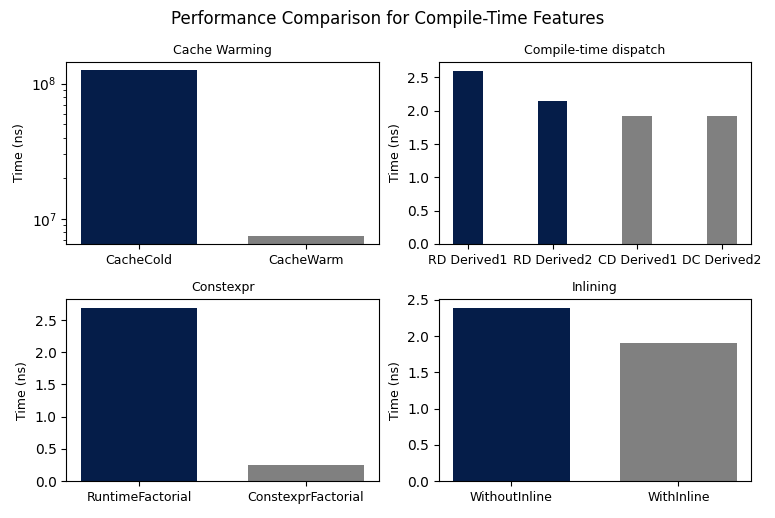}
      \caption{Performance Comparison for Compile-Time Features.}
      \label{fig:sample-image}
\end{figure}

\subsection{Optimisation Techniques}
\subsubsection*{Loop unrolling}
Loop unrolling is a technique used in computer programming to optimise the execution of loops by reducing or eliminating the overhead associated with loop control \cite{fahringer}. It's a form of trade-off between processing speed and program size. The technique involves duplicating the contents of the loop so that there are fewer iterations, which decreases the number of conditional jumps the CPU has to make. Each iteration of the loop performs more work, allowing the program to execute faster. In the given test, two functions were benchmarked: one with a regular loop (\texttt{BM\_LoopWithoutUnrolling}) and one with loop unrolling (\texttt{BM\_LoopWithUnrolling}).

\begin{lstlisting}[style=customc++, language=C++, caption={Loop unrolling Google Benchmark}]

// Benchmark function without loop unrolling
static void BM_LoopWithoutUnrolling(benchmark::State& state) {
  for (auto _ : state) {
    int result = 0;
    for (int i = 0; i < state.range(0); ++i) {
      result += i;
    }
    benchmark::DoNotOptimise(result);
  }
}
BENCHMARK(BM_LoopWithoutUnrolling)->Arg(1000);

// Benchmark function with loop unrolling
static void BM_LoopWithUnrolling(benchmark::State& state) {
  for (auto _ : state) {
    int result = 0;
    for (int i = 0; i < state.range(0); i += 4) {
      result += i + (i + 1) + (i + 2) + (i + 3);
    }
    benchmark::DoNotOptimise(result);
  }
}
BENCHMARK(BM_LoopWithUnrolling)->Arg(1000);

\end{lstlisting}

The time taken for the unrolled loop was significantly less than the standard loop. The standard loop took around 4539 ns to execute, while the unrolled loop took only 1260 ns, illustrating that the loop unrolling resulted in an improvement by \(72.24\%\). However, there are some challenges and considerations. Loop unrolling can increase the binary size, potentially causing more instruction cache misses. The performance benefits can also experience diminishing returns if a loop is excessively unrolled. Furthermore, loops that are heavily dependent on memory access speeds might not benefit significantly from unrolling. Therefore, while loop unrolling can be advantageous, its efficacy depends on specific loop characteristics, hardware considerations, and surrounding code, necessitating careful analysis and profiling.

\subsubsection*{Short-circuiting}
Short-circuiting is a logical operation in programming where the evaluation of boo\-lean expressions stops as soon as the final result is determined \cite{14}. For example, in a logical OR operation (\( A \lor B \)), if \( A \) is true, there's no need to evaluate \( B \) because the overall expression is already determined to be true, regardless of \( B \)'s value. This concept can significantly improve performance by avoiding unnecessary computations.

The given code involves benchmarking tests comparing the performance of short-circuiting and non-short-circuiting operations. Two functions, \texttt{BM\_NoShortCi\-rcuit} and \texttt{BM\_ShortCircuit}, represent these two operations. The function \texttt{BM\_NoSh\-ortCircuit} does not use short-circuiting, while \texttt{BM\_ShortCircuit} does. The benchmark results show that \texttt{BM\_ShortCircuit} (which uses short-circuiting) consistently performs faster than \texttt{BM\_NoShortCircuit} (which does not use short-circuiting). For instance, at the 8 iterations level, \texttt{BM\_ShortCircuit} operates roughly two times faster than \texttt{BM\_NoShortCircuit}. Even at higher iteration levels, such as 8192, \texttt{BM\_Short\-Circuit} demonstrates superior performance, completing in roughly half the time as \texttt{BM\_NoSh\-ortCircuit}. These results highlight the computational efficiency gained through short-circuiting, especially when dealing with expensive computations.

Based on the benchmark results provided, the short-circuiting operation (\texttt{BM\_S\-hortCircuit}) offers significant performance improvements compared to the non-short-circuiting operation (\texttt{BM\_NoShortCircuit}). The results can be seen in Table 2.

\begin{table}[H]
\centering
\begin{tabular}{lrrr}
\toprule
\textbf{Iterations} & \textbf{NoShortCircuit (ns)} & \textbf{ShortCircuit (ns)} & \textbf{Improvement (\%)} \\
\midrule
8     & 11,264,440 & 5,638,579  & 49.96 \\
64    & 89,993,360 & 45,423,755 & 49.53 \\
512   & 720,858,287 & 363,692,844 & 49.56 \\
4,096 & 5,823,645,596 & 2,900,137,804 & 49.78 \\
8,192 & 12,400,000,000 & 5,844,914,620 & 52.83 \\
\bottomrule
\end{tabular}
\caption{Comparison between BM\_NoShortCircuit and BM\_ShortCircuit}
\label{tab:comparison}
\end{table}

Therefore, across all these iterations, short-circuiting consistently results in approximately a 50\% improvement in speed.

\subsubsection*{Slowpath removal}

Separating slowpath code from the hotpath can significantly enhance latency in code execution \cite{8}. In computing, a hotpath refers to the segment of code that is frequently executed, forming the critical path in terms of performance. Conversely, slowpath code pertains to less frequently accessed, heavier operations, like error handling, logging, or non-critical routines. The principle suggests to modularize and isolate such slowpath code from the hotpath. This ensures that the processor's instruction cache, which is limited in capacity, remains primarily dedicated to serving the hotpath, thereby reducing unnecessary cache pollution and context switches.  The first code snippet represents a less optimised design where error handling is directly included in the decision structure. The second code snippet proposes a better design where a function, \texttt{HandleError}, which is marked as non-inline, encapsulates the slowpath operations. This approach keeps the slowpath code away from the hotpath, reducing instruction cache load and improving overall performance.

\begin{minipage}{.45\textwidth}
    \begin{lstlisting}[language=C++, title=Snippet 1: Bad design]
if (checkForErrorA())
    handleErrorA();
else if (checkForErrorB())
    handleErrorB();
else if (checkForErrorC())
    handleErrorC();
else
    sendOrderToExchange();
    \end{lstlisting}
  \end{minipage}\hfill%
  \begin{minipage}{.45\textwidth}
    \begin{lstlisting}[language=C++, title=Snippet 2: Good design]
int64_t errorFlags;
...
if (!errorFlags)
    sendOrderToExchange();
else
    HandleError(errorFlags);
\end{lstlisting} 
\end{minipage}\\

The benchmark to test this programming strategy implemented two benchmark scenarios: \texttt{SlowpathNotRemoved} and \texttt{SlowpathRemoved}. In both scenarios, the hotpath is hit \(90\%\) of the time, and the \texttt{slowpath} \(10\%\) of the time. In the \texttt{SlowpathNotRemoved} scenario, slowpath operations (i.e., \texttt{sendToClient}, \texttt{remove\-FromMap}, \texttt{logMessage}) are executed directly. Conversely, in the \texttt{SlowpathRemoved} scenario, these operations are encapsulated inside the \texttt{HandleError} function, which is explicitly marked as non-inline to ensure it does not bloat the hotpath code.

\begin{lstlisting}[style=customc++, language=C++, caption={Slowpath removed Google Benchmark}]
static void SlowpathNotRemoved(benchmark::State& state) {
  int counter = 0;
  for (auto _ : state) {
    okay = (++counter % 10 != 0);
    if (okay) {
      // Hotpath.
    } else {
      // Slow-path
      sendToClient("Order failed");
      removeFromMap(1, orders);
      logMessage("Order failed");
    }
  }
}

static void SlowpathRemoved(benchmark::State& state) {
  int counter = 0;
  for (auto _ : state) {
    okay = (++counter % 10 != 0);
    if (okay) {
      // Hotpath.
    } else {
      // Slow-path
      HandleError();
    }
  }
}
\end{lstlisting}

Based on the benchmark results, the optimised scenario (\texttt{SlowpathRemoved}) exhibits a significant performance improvement over the less optimised one (\texttt{Sl\-owpathNotRemoved}). Specifically, the execution time reduces from \(28074\) nanoseconds to \(24755\) nanoseconds, indicating approximately a \(12\%\) performance improvement. This boost in speed arises primarily due to the efficient use of the processor's instruction cache. In the optimised scenario, slowpath code is modularized into the non-inline function \texttt{HandleError}, thus reducing the footprint on the instruction cache, especially during frequent hotpath execution. Consequently, cache hits are maximized, resulting in less time spent on fetching instructions from memory. This demonstrates the effectiveness of separating slowpath code from the hotpath as a design strategy in improving latency.

\subsubsection*{Branch reduction}
Branch prediction has become a cornerstone of contemporary Central Processing Units (CPUs), where it functions to anticipate the path that will be followed in a branching operation. When this prediction is accurate, it allows for the prefetching of instructions from memory, leading to seamless and efficient execution. However, a mispredicted branch results in a penalty, causing the CPU to stall until the correct instructions are fetched, thus hampering execution speed. This context underscores the potential value of implementing compile-time branch prediction hints, such as \texttt{\_\_builtin\_expect} in GCC. These hints have the potential to guide the compiler in generating more optimal code structures. Furthermore, another essential approach to minimising latency includes isolating error handling from the frequently executed parts of the code (hotpath), hence reducing the number of branches, and consequently, the opportunities for misprediction. This is branch reduction \cite{8}.

In a conventional programming approach, the handling of errors and the execution of the main function (often referred to as the ``\texttt{hotpath}") are often intertwined in a series of \texttt{if-else} statements. For instance, as demonstrated in the first code snippet, each potential error is checked in sequence; only when no errors are detected does the program proceed to execute the main function. This structure is quite common, yet it presents opportunities for branch misprediction, as the direction of execution could change at each conditional check.

In contrast, the alternative approach, illustrated in the second code snippet, adopts a distinct strategy that fundamentally separates error handling from the main function execution. Here, an integer variable (in this case, \texttt{errorFlags}) is utilised to represent the presence of different errors. If an error is detected, it is handled by a dedicated function \texttt{HandleError(errorFlags)}, which uses the value of \texttt{errorFlags} to decide the specific error handling routine. If no errors are detected, the program proceeds directly to the main function, bypassing all the conditional checks present in the conventional approach. This method reduces the chances of branch misprediction and could potentially lead to improved program execution speed. This hypothesis forms the basis for the empirical analysis conducted to evaluate the impact of this approach on program execution latency.

\begin{minipage}{.45\textwidth}
    \begin{lstlisting}[language=C++, title=Snippet 1: Bad design]
if (checkForErrorA())
    handleErrorA();
else if (checkForErrorB())
    handleErrorB();
else if (checkForErrorC())
    handleErrorC();
else
    executeHotpath();
    \end{lstlisting}
  \end{minipage}\hfill%
  \begin{minipage}{.45\textwidth}
    \begin{lstlisting}[language=C++, title=Snippet 2: Good design]
uint32_t errorFlags;
...
if (errorFlags)
HandleError(errorFlags)
else
{
... hotpath
}
\end{lstlisting}
\end{minipage}

The benchmark for branch reduction consisted of two distinct code structures, one implementing the conventional branching scheme and the other utilising a reduced branching technique. In the conventional scheme, error handling was interlaced with the critical execution path, whereas, in the latter, a bitwise operation was employed to check for errors, thereby separating it from the hotpath.

 The results of the benchmark revealed performance improvements with the adoption of the reduced branching technique. The conventional branching approach resulted in an average execution time of 7.35 nanoseconds per iteration. In contrast, the reduced branching technique yielded an average time of 4.68 nanoseconds per iteration, marking an approximate speed increase of 36\%. Thus, the findings corroborate the thesis that by minimising unnecessary branches and avoiding potential branch mispredictions, substantial reductions in program latency can be achieved.
 
\subsubsection*{Prefetching}
Prefetching is a technique used by computer processors to boost execution performance by fetching data and instructions from the main memory to the cache before it is actually needed for execution \cite{8, 14}. It anticipates the data needed ahead of time, and therefore, when the processor needs this data, it is readily available from the cache, rather than having to fetch it from the slower main memory. Prefetching reduces latency as it minimises the time spent waiting for data fetch operations, allowing a more efficient use of the processor's execution capabilities. The technique is beneficial in scenarios where data access patterns are predictable, such as traversing arrays, processing matrices, or accessing data structures in a sequential manner. In these scenarios, prefetching can significantly lower the latency, resulting in faster code.

The benchmark to test this programming strategy included two functions: \texttt{NoPrefetch} and \texttt{WithPrefetch}, both of which sum up the elements of a very large vector. However, the \texttt{WithPrefetch} function also includes the \texttt{\_\_builtin\_prefetch} command to prefetch data. Analyzing the benchmark results, it's evident that pre\-fetching has improved the execution speed of the \texttt{WithPrefetch} function by approximately 23.5\% (8235924 ns for \texttt{NoPrefetch} vs. 6301400 ns for \texttt{WithPrefetch}). This improvement is due to the prefetching technique used in the \texttt{WithPrefetch} function, which preemptively loads the data into the cache before it is needed. This reduces the latency associated with fetching data from the main memory. The processor, in the \texttt{WithPrefetch} function, experiences fewer cache misses and thus spends less time waiting for data from the main memory, which results in an overall speedup of the operation. It should be noted that the actual improvement may depend on several factors such as the processor's architecture, the size of the data set, and the predictability of the data access pattern.

\begin{figure}[H]
      \centering
      \includegraphics[width=0.8\textwidth]{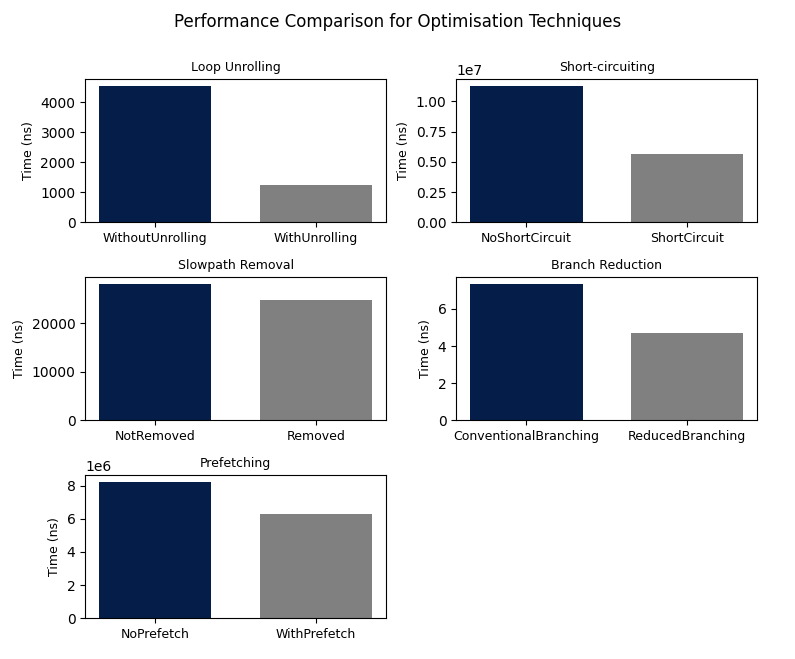}
      \caption{Performance Comparison for Optimisation Techniques.}
      \label{fig:sample-image}
\end{figure}

\subsection{Data Handling}
\subsubsection*{Signed vs unsigned comparisons}
In computer programming, the terms `signed' and `unsigned' refer to whether or not a binary number is capable of representing negative values \cite{14}. `Signed' means that a binary number can represent both positive and negative numbers, whereas `unsigned' means that a binary number can only represent positive numbers. This distinction plays a role in various aspects of programming, including arithmetic operations, comparisons, and loop control as we have in this example.

The benchmark test was conducted to investigate the speed difference between using a signed integer versus an unsigned integer in a specific scenario -- a loop operation where a value of \( i \) is incremented by 1 for 10 iterations. Two functions, \texttt{f\_signed} and \texttt{f\_unsigned}, were created, one using a signed integer for the loop variable and the other using an unsigned integer. These functions were then benchmarked using Google's Benchmark library, with the \texttt{rand()} function used to generate a random input value \( i \) for each function. The objective of the test was to compare the execution speed of these two versions of the function, and thereby ascertain whether using a signed or unsigned integer had any impact on performance.

\begin{figure}[ht]
\begin{minipage}{0.45\textwidth}

\begin{lstlisting}
f(int):
        mov     eax, 10
        ret
main:
        sub     rsp, 8
        call    rand
        mov     eax, 10
        add     rsp, 8
        ret
\end{lstlisting}
\end{minipage}\hfill
\begin{minipage}{0.45\textwidth}

\begin{lstlisting}
f(int):
        lea     eax, [rdi+10]
        cmp     edi, eax
        sbb     eax, eax
        and     eax, 10
        ret
main:
        sub     rsp, 8
        call    rand
        lea     edx, [rax+10]
        cmp     eax, edx
        sbb     eax, eax
        add     rsp, 8
        and     eax, 10
        ret
\end{lstlisting}
\end{minipage}
\caption{Signed (left) vs. Unsigned (right) assembly code}
\end{figure}
According to the benchmark results, the function using the signed integer (\texttt{BM\_Signed}) had an average execution time of 0.282 nanoseconds. The function using the unsigned integer (\texttt{BM\_Unsigned}), on the other hand, had an average execution time of 0.321 nanoseconds. This indicates that, based on these specific benchmark results, using a signed integer instead of an unsigned integer led to a roughly 12.15\% speed improvement.

The reason for this difference lies in the assembly code. The additional instructions in the assembly code generated for the \texttt{f\_unsigned} function compared to the \texttt{f\_signed} function are due to how the compiler is handling the potential for overflow with the unsigned integer. In the \texttt{f\_unsigned} function, when \(i\) is close to the maximum value that an integer can hold, adding 10 to it can result in an overflow. An overflow with an unsigned integer is defined in C++ and causes the integer to wrap around and become a small value. Thus, the loop \texttt{for (unsigned k = i; k < i + 10; ++k)} may not behave as expected and could become an infinite loop. The \texttt{cmp}, \texttt{sbb}, and bitwise \texttt{AND} operations are part of the compiler's attempt to handle this scenario correctly. The \texttt{lea} instruction is used to compute the effective address \(i + 10\) once and use it in the loop. The \texttt{cmp} and \texttt{sbb} instructions are used to check whether \(i + 10\) has overflowed. The result of \texttt{sbb} is 0 if there was no overflow and -1 if there was. The \texttt{and} operation with 10 is used to get 0 if there was an overflow (an infinite loop is avoided) and 10 otherwise.

In contrast, for the \texttt{f\_signed} function, when \(i\) is large, adding 10 to it causes an overflow. In the case of signed integers, such an overflow is technically undefined behavior in C++, but many compilers handle this by allowing the value to wrap around to a large negative number. The assembly code for \texttt{f\_signed} reflects this: it moves the constant 10 into the \texttt{eax} register and returns this value, without any concern for handling overflow scenarios. Therefore, these extra instructions in the unsigned case are present to handle the overflow case correctly and safely, and this additional work is what causes the \texttt{f\_unsigned} function to run slower than the \texttt{f\_signed} function.

\subsubsection*{Mixing data types}
%Mixing floats and doubles can lead to minor runtime cost due to the necessity of type conversions \cite{14}. When a computation involves both float and double types, implicit conversions are required. For example, in the operation float\_var * 1.23, where float\_var is a float, the double literal 1.23 has higher precision. Hence, float\_var needs to be temporarily promoted to double for the computation, resulting in a double result. Then, the double result needs to be demoted back to float to match the original variable type, which adds computational overhead.

%The process of conversion between float and double is not trivial, particularly when such operations are executed in a loop or a function that is called frequently. The processing power needed for these conversions may seem negligible in a single operation, but in high-performance computing scenarios where these conversions are done millions or billions of times, the added runtime cost can significantly impact the performance of the software. The loss of speed is a trade-off for the increase in precision when promoting from float to double. However, when the value is demoted back to float, there could also be a loss of accuracy, introducing potential issues in the precision of the computations.

Mixing floats and doubles can result in a minor runtime cost due to the need for type conversions \cite{14}. When a computation involves both \texttt{float} and \texttt{double} types, implicit conversions are required. For instance, consider the operation \(\texttt{float\_var} \times 1.23\), where \(\texttt{float\_var}\) is a \texttt{float}. The \texttt{double} literal \(1.23\) has a higher precision, which necessitates the temporary promotion of \(\texttt{float\_var}\) to \texttt{double} for the computation. The result is then a \texttt{double}, which must be demoted back to \texttt{float} to match the original variable type, adding computational overhead.

The conversion process between \texttt{float} and \texttt{double} is far from trivial, especially when such operations occur within loops or are part of frequently called functions. While the computational cost of these conversions may appear negligible for isolated operations, they can become significant in high-performance computing scenarios where such conversions happen millions or billions of times. The speed loss is the trade-off for the increased precision when promoting from \texttt{float} to \texttt{double}. However, demoting the value back to \texttt{float} could also introduce a loss of accuracy, thereby posing potential precision-related issues.

The benchmark analysis was conducted to measure and compare the performance of two C++ functions, namely \texttt{BM\_MixedFloatDouble} and \texttt{BM\_Unmixed\-FloatDouble}. These functions differed in their approach to handling floating-point number computations. In \texttt{BM\_MixedFloatDouble}, a floating-point operation involving both \texttt{float} and \texttt{double} types was performed. Within the function, two \texttt{float} variables \(a\) and \(b\) were declared, with \(b\) being assigned a random value. \(b\) was then multiplied by the \texttt{double} literal \(1.23\), and the resulting product was stored in \(a\). This multiplication required an implicit conversion from \texttt{float} to \texttt{double} and then back to \texttt{float}, thereby accommodating the higher precision of the \texttt{double} literal. In contrast, \texttt{BM\_UnmixedFloatDouble} involved computations exclusively with \texttt{float} types. Similar to the first function, it declared \(a\) and \(b\) as \texttt{float} variables and assigned \(b\) a random value. However, the multiplication used a \texttt{float} literal \(1.23f\), thus eliminating the need for any conversions between \texttt{float} and \texttt{double}.

The results of the benchmarking were revealing. The \texttt{BM\_MixedFloatDouble} function took approximately \(21.6\) nanoseconds to execute, while the \texttt{BM\_Unmixed\-FloatDouble} function required around \(14.2\) nanoseconds. The calculations showed that the version using only \texttt{float} computations and no implicit conversions, \texttt{BM\_Un\-mixedFloatDouble}, was approximately \(52\%\) faster than the mixed version. This significant speed improvement supported the initial hypothesis, indicating that the overhead of converting between \texttt{float} and \texttt{double} can indeed introduce a non-negligible performance cost, particularly in time-sensitive or high-performance computing applications.

\begin{figure}[H]
      \centering
      \includegraphics[width=0.7\textwidth]{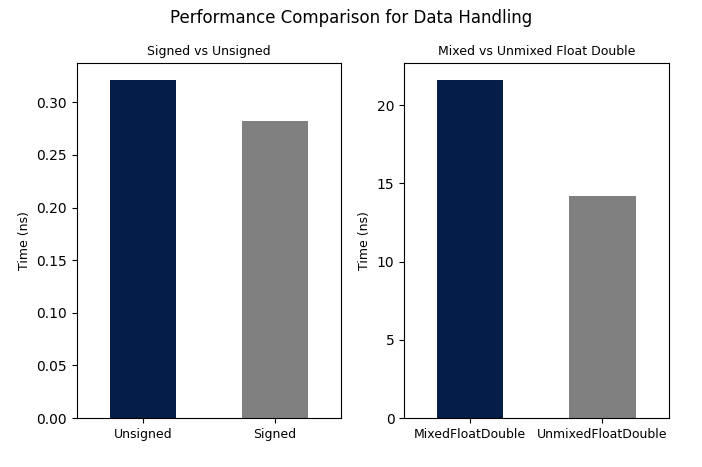}
      \caption{Performance Comparison for Data Handling.}
      \label{fig:sample-image}
\end{figure}

\subsection{Concurrency}
\subsubsection*{SIMD Instructions}
%SIMD, or Single Instruction Multiple Data, is a class of parallel computing architectures where a single operation can be performed on multiple data points simultaneously \cite{flynn}. Modern CPUs have built-in support for SIMD operations via specific instructions that can process data in parallel \cite{asaduzzaman}. For example, in this code, SSE2 (Streaming SIMD Extensions 2) is used, which is a SIMD instruction set extension for the x86 architecture developed by Intel. It allows for parallel processing of packed floating-point numbers in 128-bit registers, significantly speeding up computations on large arrays of data.

%The benchmark test compares the performance of two functions: add\_arrays and add\_arrays\_simd. The former is a regular function that adds elements from two arrays one by one in a loop. The latter, however, utilises SIMD instructions to perform the addition on four elements at a time. The results show that the SIMD-optimised function (BM\_ArrayAddition\_SIMD) operates significantly faster, taking an average of 10,929 nanoseconds per operation, compared to the non-optimised function (BM\_ArrayAddition), which takes 21,447 nanoseconds per operation. This corresponds to a roughly 49\% reduction in operation time. The primary reason for this improvement is that SIMD allows multiple data points to be processed simultaneously, thus greatly increasing the data throughput and reducing latency, particularly beneficial when operating on large data sets or arrays.

SIMD, or Single Instruction Multiple Data, is a category of parallel computing architectures where a single instruction can act upon multiple data points simultaneously \cite{flynn}. Modern CPUs offer built-in support for SIMD operations through specific instruction sets designed for parallel data processing \cite{asaduzzaman}. For instance, the code in this example utilizes SSE2 (Streaming SIMD Extensions 2), a SIMD instruction set extension for the x86 architecture developed by Intel. SSE2 allows for the parallel processing of packed floating-point numbers stored in 128-bit registers, thereby substantially accelerating calculations on extensive data arrays.

The benchmark test focuses on comparing the performance of two functions: \texttt{add\_arrays} and \texttt{add\_arrays\_simd}. The former function performs the addition of array elements in a sequential manner within a loop. The latter, on the other hand, employs SIMD instructions to execute the addition on four elements simultaneously. The benchmark results indicate a significant speed advantage for the SIMD-optimized function (\texttt{BM\_ArrayAddition\_SIMD}), which has an average execution time of \(10,929\) nanoseconds per operation. In contrast, the non-optimized function (\texttt{BM\_ArrayAddition}) takes approximately \(21,447\) nanoseconds per operation. This translates to a reduction in operation time by roughly \(49\%\). The primary contributor to this speed-up is the SIMD architecture's capability to process multiple data points concurrently, leading to increased data throughput and reduced latency, features especially beneficial for operations involving large data sets or arrays.

\subsubsection*{Lock-Free Programming}

Lock-free programming is a concurrent programming paradigm that centers around the construction of multi-threaded algorithms which, unlike their traditional counterparts, do not employ the usage of mutual exclusion mechanisms, such as locks, to arbitrate access to shared resources \cite{herlihy}. The key concept that underpins this approach is the use of atomic operations which, in the context of multi-threaded programming, are operations that either fully complete or do not execute at all without the interference of other threads. The ability to guarantee such properties, while simultaneously maintaining data integrity across threads, eliminates the issues of deadlocks, livelocks, and other synchronization-related bottlenecks that are prevalent in lock-based designs, hence improving system-wide latency and throughput \cite{fraser}. Lock-free programming achieves its objectives by employing low-level synchronization primitives, such as Compare-and-Swap (CAS) or Load-Link/Store-Conditional (LL/SC), which are typically provided by modern processor architectures \cite{anderson}. These operations are the cornerstone of lock-free data structures and are used to resolve conflicting access to shared data. For instance, the CAS operation, given three operands - memory location, expected old value, and new value - atomically alters the memory location to the new value only if the current value at that location matches the expected old value. If the current value does not match, the operation is retried, usually in a loop until it succeeds. This operation is central to the implementation of many lock-free algorithms. While the latency and throughput improvements provided by lock-free programming can be quite significant, it's important to underscore that lock-free programming is more complex than its lock-based counterparts and is thus more prone to bugs, particularly if misused \cite{mckenney}. Care must be taken to correctly design and test lock-free algorithms as the failure modes can be non-obvious and difficult to debug. In addition, not all problems lend themselves well to a lock-free solution, and so, it's important to use these techniques judiciously and only when necessary.

The benchmarking code conducts an empirical evaluation comparing the performance of two distinct methods employed to increment a counter within the context of a multi-threaded environment: one facilitated through an atomic operation (\texttt{increment\_atomic}) and another orchestrated through the deployment of a mutex (\texttt{increment\_locked}). The primary technology utilised in executing the benchmark is Google Benchmark, a library renowned for its microbenchmarking capabilities. Each function designed for the benchmarking exercise is registered with Google Benchmark via the \texttt{BENCHMARK} macro, with each being subsequently executed with an argument quantified by \texttt{->Arg(10000)}. This argument effectively delineates the frequency of increment operations performed within each invocation of the corresponding function. The function \texttt{increment\_atomic} leverages an atomic integer (\texttt{atomic\_counter}) as the cornerstone of the counter implementation. The increment operation (\texttt{++}) executed on the atomic integer is certified to be atomic, a quality that guarantees the operation will either be executed in full or not at all, irrespective of the potential simultaneous calls from a multitude of threads. This level of thread-safety is accomplished without the need for additional synchronization mechanisms. On the contrary, the function \texttt{increment\_locked} harnesses a rudimentary integer (\texttt{locked\_counter}) in tandem with a mutex (\texttt{counter\_mutex}) to enforce thread safety. Each increment operation is encapsulated within a \texttt{std::lock\_guard} block, which locks the mutex at the commencement of the block and unlocks it at its conclusion (precisely when the \texttt{lock\_guard} is destructed). This design ensures exclusive access, wherein only one thread is capable of incrementing the counter at any given moment. The most glaring discrepancy between these two methodologies lies in the necessity, or lack thereof, of locking. The process of locking can be computationally expensive, necessitating system calls and occasionally leading to the suspension of the calling thread if the lock is presently held by an alternate thread. In contrast, atomic operations are typically implemented using CPU instructions and do not demand context switches, thus conferring upon them a superior speed profile. As such, it is anticipated that the code will demonstrate a faster performance profile for the incrementation of the atomic counter as opposed to the locked counter, thereby illustrating how atomic operations can yield performance improvements in specific multi-threaded contexts.

The empirical results from the benchmarking exercise provide compelling insights into the temporal efficiencies of atomic versus mutex-based increment operations. The atomic incrementation (\texttt{BM\_Atomic/10000}) registers an execution time of approximately \(65,369 \, \text{ns}\), while the mutex-based incrementation (\texttt{BM\_Mutex/10000}) records a significantly higher execution time of approximately \(175,904 \, \text{ns}\). Given these observations, it becomes apparent that the atomic approach markedly outperforms the mutex-based method under the tested conditions. Specifically, the atomic operation resulted in a performance improvement of approximately \(63\%\) when compared to the mutex-based method. This data reinforces the theoretical notion that lock-free programming, which includes atomic operations, can offer substantial performance benefits, especially in scenarios involving high contention. It's important to underline that while these results are significant, the actual performance improvement may vary based on numerous factors, including hardware specifications, number of threads, and the workload's contention level.

\begin{figure}[H]
      \centering
      \includegraphics[width=0.8\textwidth]{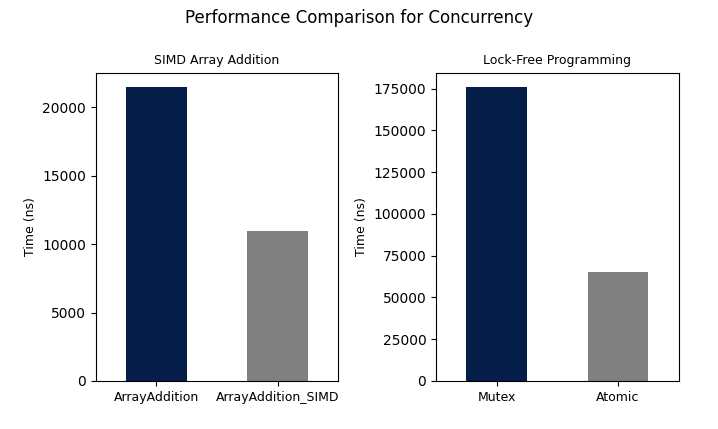}
      \caption{Performance Comparison for Concurrency.}
      \label{fig:sample-image}
\end{figure}

\subsection{System Programming}
\subsubsection*{Kernel Bypass}
Kernel bypass is a technique that can significantly enhance the latency and throughput of a computer system, primarily by increasing speed. This is not a technique which was benchmarked in this work, however it is still a key topic which is covered to provide readers with a comprehensive arsenal of programming techniques. In a typical operating system architecture, network I/O (input/output) operations must pass through the operating system's kernel. The kernel is the core component of an OS, managing tasks such as system calls, memory, and process scheduling. When a network packet is received, it's processed by the network card, then passed to the kernel, which then copies it to user space for the application to process. This process involves multiple context switches between user space and kernel space, leading to relatively high latency due to overheads, such as the CPU cycles required for these switches, and memory copying overheads.

Kernel bypass mitigates these latency issues by facilitating direct communication between user applications and the network interface card (NIC). This is done by removing the kernel's involvement in the data path for sending and receiving packets, thus 'bypassing' the kernel. By leveraging this approach, the need for data copying and context switching is greatly reduced. Applications can write data directly to the network card's buffers and read data straight from these buffers. In essence, the I/O operations are performed in user space instead of the kernel space, which saves significant amounts of time and thereby reduces latency. This is how kernel bypass can improve the speed of operations \cite{Qazi}.

It's important to note, however, that kernel bypass requires specialized network cards and drivers that provide such capabilities. The most widely used software that provides kernel bypass functionality includes Solarflare's OpenOnload, Mellanox's VMA, and Intel's DPDK (Data Plane Development Kit). Barbosa discussed the use of kernel bypass with DPDK and states that DPDK can achieve a 7-fold decrease in latency when compared to a traditional Linux networking stack \cite{Barbosa}. These tools allow applications to interact with the hardware directly, therefore reducing overheads and increasing speed. Additionally, they require careful tuning and management, as bypassing the kernel's traditional security and isolation mechanisms can increase the risk of system instability or security vulnerabilities if not managed correctly. Despite these challenges, kernel bypass offers a compelling advantage in high-frequency trading, real-time analytics, high-performance computing, and other latency-sensitive applications \cite{ThomasR}.

\subsection{Results}

The benchmarked results of the Low-Latency Programming Repository can be seen in Figure 11. Cache warming and constexpr exhibited the most dramatic effects, boosting speed by approximately 90\% each. Loop unrolling also made a considerable impact, achieving a 72\% speed increase. Other techniques like Lock-Free Programming and Short-circuiting showed strong results, with speed improvements of 63\% and 50\% respectively. SIMD instructions were not far behind at 49\%, while mixing data types yielded a 52\% improvement. On the other hand, some techniques showed more moderate improvements: Compile-time dispatch at 26\%, Inlining at 20.5\%, Prefetching at 23.5\%, and Branch reduction at 36\%. The least impactful optimisations were slowpath removal and signed vs unsigned comparisons, which only achieved 12\% and 12.15\% improvements, respectively. Overall, these findings provide insights into the potential for enhancing computational efficiency using different optimisation techniques.

\begin{figure}[H]
      \centering
      \includegraphics[width=1\textwidth]{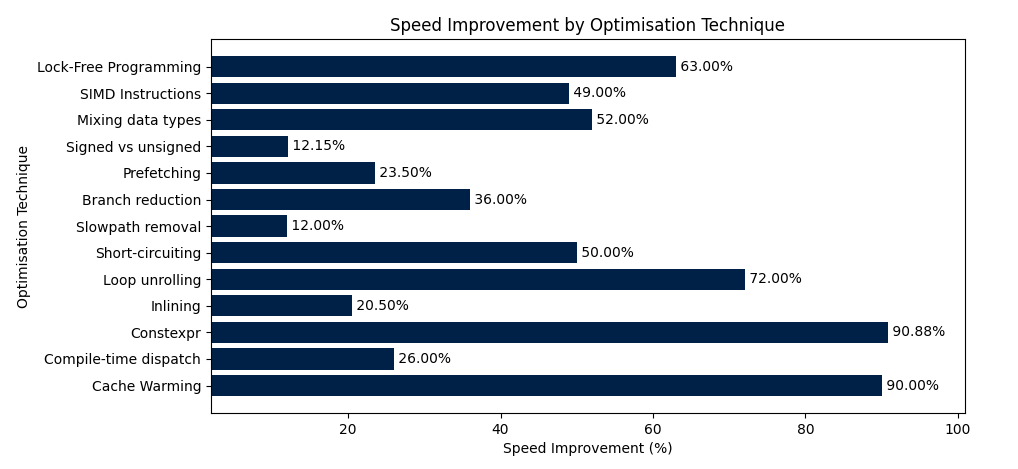}
      \caption{Speed Improvement by Optimisation Technique.}
      \label{fig:sample-image}
\end{figure}

\section{Pairs Trading}
\subsection{Introduction to strategy}
The design patterns which were tested in isolation were combined in a statistical arbitrage pairs trading strategy to highlight a real life implementation of the discussed ideologies above. Pairs trading, also referred to as relative value arbitrage, constitutes a market-neutral quantitative trading strategy that leverages statistical and computational approaches to pinpoint asset pairs that display historical price correlations \cite{gatev, elliott, chan_quantitative_trading}. This strategy creates a unique opportunity for investors to capitalize on the dynamics of price behavior, particularly on the divergences and convergences that occur in the relationship between the pair of assets \cite{elliott, chan_quantitative_trading}. The inception of pairs trading can be traced back to the late 1980s, attributed to the pioneering efforts of a quantitative group at Morgan Stanley led by Nunzio Tartaglia \cite{gatev, chan_algorithmic_trading}. The strategy was built on the observation that certain pairs of securities demonstrated a predictable relationship of price behavior over time, which laid the groundwork for this trading approach.

The fundamental principle that underpins pairs trading resides in the mean-reverting nature of the price difference or spread between the identified correlated pair of assets \cite{avellaneda, jaeger}. Simply put, mean reversion is the financial theory suggesting that the prices of assets, over time, tend to move towards their average, or 'mean'. Hence, in pairs trading, the investor capitalizes on the assumption that the spread between the pair will revert to its historical mean value. In practice, this typically manifests in the form of a long position in the underpriced asset, paired with a short position in the overpriced one, when a significant divergence in their price relation occurs. This dual stance is believed to hedge against market risks, rendering the strategy market-neutral as it seeks to profit from the relative price movement of the two assets, regardless of the direction of the broader market trends \cite{chan_algorithmic_trading, chan_quantitative_trading}.

The selection of pairs often involves shares from the same industry or sector, to mitigate idiosyncratic risk associated with individual companies. The process of identification and selection is primarily guided by econometric methods, such as cointegration tests or correlation analysis \cite{chan_machine_trading, chan_algorithmic_trading, chan_quantitative_trading}. Cointegration tests aid in confirming whether the pair of assets maintains a stable long-run relationship, despite short-term fluctuations, thus ensuring the stability of the spread over time. On the other hand, correlation analysis helps determine the extent to which the asset pair moves in tandem. These paired assets should ideally have a high degree of price correlation and a stable long-term equilibrium, ensuring the resilience of the strategy in diverse market conditions.

The overall execution of pairs trading extends beyond the algorithmic aspects, requiring robust risk management procedures to account for market volatilities and other unpredictable elements. This aspect, albeit not covered in this particular study due to its focus on the strategic concept, is paramount to the successful application of the pairs trading strategy. In summary, pairs trading is an innovative investment strategy predicated on the quantitative analysis of asset prices. By leveraging the mean-reverting nature of price spreads between correlated assets, it provides a market-neutral approach to investment, poised to generate returns irrespective of market movements, provided that rigorous selection criteria and risk management practices are maintained.

\subsection{Cointegration}
Cointegration is a critical concept in time-series analysis, particularly in the realm of econometrics, where it addresses the long-term relationships between non-stationary variables \cite{engle_granger, chan_algorithmic_trading}. When two or more non-stationary time-series exhibit a common stochastic trend, they are said to be cointegrated \cite{engle_granger}. Stationarity refers to the condition where a time-series' statistical properties do not change over time, and non-stationarity implies the opposite \cite{engle_granger}. The primary motivation behind cointegration analysis arises from the need to distinguish spurious relationships from genuine long-term associations between variables. Mathematically, let us consider two non-stationary time-series, $Y_t$ and $X_t$, which can be expressed as:

\begin{equation}
Y_t = \rho Y_{t-1} + \epsilon_{Y_t}
\end{equation}

\begin{equation}
X_t = \beta X_{t-1} + \epsilon_{X_t}
\end{equation}

where $\rho$ and $\beta$ are coefficients, and $\epsilon_{Y_t}$ and $\epsilon_{X_t}$ are white noise error terms. If $Y_t$ and $X_t$ are both non-stationary but their linear combination, say $Z_t=Y_t - \gamma X_t$, is stationary, then $Y_t$ and $X_t$ are cointegrated. This condition implies that any deviation from their long-term equilibrium relationship, represented by $\gamma$, will be mean-reverting. The statistical procedure to test for cointegration involves several steps. Initially, unit root tests like the Augmented Dickey-Fuller (ADF) test are employed to determine whether the individual time-series are non-stationary \cite{engle_granger, dickey_fuller}. Next, the residuals from the regression of one time-series on the other, after checking for the order of integration (usually using ADF or Phillips-Perron tests), are tested for stationarity using tests like the ADF test again. If the residuals are stationary, it confirms the presence of cointegration between the original time-series. Cointegration analysis is helpful in modeling long-term relationships between variables, such as in financial markets, macroeconomics, and other fields with interrelated time-series data. Multiple US equities were tested for cointegration over five year period with daily closing price as the intervals. Goldman Sachs Group Inc (US:GS) and Morgan Stanley (US:MS) showed strong cointegration. The data for this analysis was obtained from Yahoo! Finance containing the relevant adjusted closing prices. To conduct the cointegration test, the Engle-Granger two-step methodology was implemented from the coint function from the statsmodels.tsa.stattools Python module. This two-step procedure is designed to identify and test the residuals from a hypothesized long-term equilibrium relationship between two time series for stationarity. The first step of the Engle-Granger method entails regressing one time series on the other using Ordinary Least Squares (OLS) to estimate a long-run equilibrium relationship, producing a series of residuals. The second step applies a unit root test, specifically the Augmented Dickey-Fuller (ADF) test, to these residuals to determine whether they are stationary. The output of this test is a test statistic (t-statistic) and an associated p-value. The t-statistic, also known as the score, is the result of the unit-root test on the residuals. A more negative t-statistic suggests that the residuals are more likely to be stationary. The p-value provides a measure of the probability that the null hypothesis of the test (no cointegration) is true. The results of your test yielded a p-value of approximately 0.0149 and a t-statistic of -3.7684. Given a commonly used significance level of 0.05, the low p-value allows us to reject the null hypothesis of no cointegration. This suggests that the adjusted closing prices of GS and MS are likely to be cointegrated over the five year period, implying that they share a long-term equilibrium relationship. In other words, deviations from this equilibrium are temporary, and the two series appear to move together over time, maintaining a long-term relationship. This statistical link is vital in various fields including but not limited to finance, for instance in pairs trading strategies. However, statistical analysis can only provide evidence and not definitive proof, hence why this is a style of statistical arbitrage trading strategies rather than pure arbitrage. Furthermore, while the Engle-Granger method is relatively simple and widely used, more complex methods such as the Johansen test are available, which can account for more than two cointegrated series and offer more robust results under certain circumstances.

\subsection{Methodology}
The algorithm begins by sourcing historical price data for Goldman Sachs and Morgan Stanley from a CSV file. In particular it used ``Adj Close" prices for generating signals. The advantage of using this over just the ``Close" price is that it factors in various elements that could influence the stock's raw closing price, such as dividends and stock splits, providing a more accurate representation of the stock's value at that given time. These adjusted close prices were subsequently stored in a vector for future computations. The cornerstone of this pairs trading strategy hinges on the accurate calculation and thorough analysis of the ``spread". In terms of pairs trading, the 'spread' embodies the difference in the adjusted close prices of the two chosen stocks. Figure 12 shows the spread between the two prices across 5 years.

\begin{figure}[H]
      \centering
      \includegraphics[width=1\textwidth]{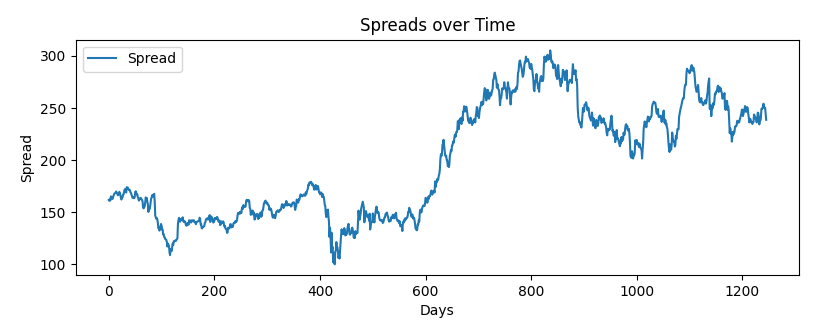}
      \caption{Spreads over time.}
      \label{fig:sample-image}
\end{figure}

With a specified window size `N' in place, the algorithm engages in the computation and examination of the spread. An essential part of understanding the behavior of the spread involves the computation of its mean and standard deviation over the window of `N' prices. The mean of the spread gives us a measure of its central tendency, representing an average state of the spread during the specified period. Concurrently, the standard deviation serves as an indicator of the spread's volatility, thereby providing insights into its risk and stability. The computation process for the mean requires the sum and sum of squares of the spread, forming a fundamental basis for the subsequent calculation of the standard deviation. Upon completion of these computations, the algorithm proceeds to analyze the current spread, derived from the difference in the current prices of the two stocks. This analysis occurs in relation to the previously computed mean and standard deviation. The algorithm employs these parameters to standardize the current spread, culminating in the derivation of a z-score. Figure 13 shows the z-scores across the 5 years. The formula for calculating the z-score is given by:

\begin{equation}
    Z = \frac{(X - \mu)}{\sigma}
\end{equation}

Where:
\begin{itemize}
    \item \( Z \) is the z-score.
    \item \( X \) is the value of the data point.
    \item \( \mu \) is the mean of the dataset.
    \item \( \sigma \) is the standard deviation of the dataset.
\end{itemize}

\begin{figure}[H]
      \centering
      \includegraphics[width=1\textwidth]{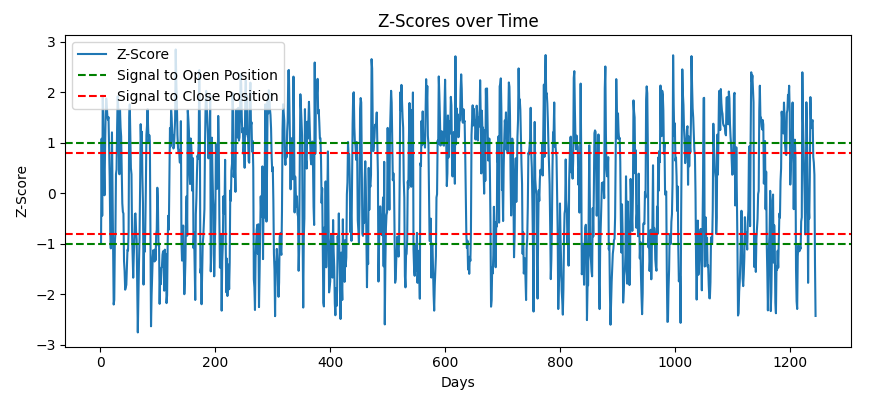}
      \caption{Z-scores over time.}
      \label{fig:sample-image}
\end{figure}

The importance of the z-score becomes evident as it is the primary catalyst for the trading signals. A z-score greater than 1.0 can be interpreted as a potential overpricing of the Goldman Sachs stock relative to Morgan Stanley. This indication prompts a short signal for the pair, where the Goldman Sachs stock would be sold, and the Morgan Stanley stock would be bought. Contrarily, a z-score less than -1.0 signifies a potential underpricing of Goldman Sachs relative to Morgan Stanley, thereby generating a long signal. Here, the Goldman Sachs stock would be bought while the Morgan Stanley stock would be sold. If the absolute value of the z-score falls below 0.8, this signifies a regression towards the mean and triggers a closure of any current positions. Any other z-score value falls into a neutral zone where no trading signal is initiated. The signals to open positions or close positions can be seen in Figure 14.

\begin{figure}[H]
      \centering
      \includegraphics[width=1\textwidth]{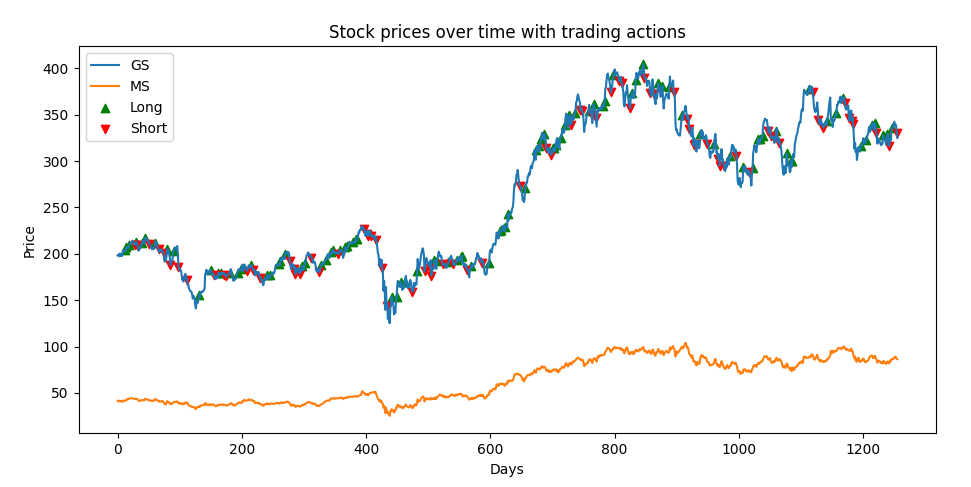}
      \caption{Stock prices over time with trading actions.}
      \label{fig:sample-image}
\end{figure}

\noindent The portfolio balance across the 5 years can be seen in Figure 15. The spikes which can be noticed in Figure 15 is due to opening long and short positions. Figure 16 shows the portfolio balance across time when no trades are open, providing a smooth time-series of only the profits. While the primary focus of this research did not revolve around evaluating the trading algorithm's effectiveness, it remains crucial to acknowledge the outcomes of the backtest. The portfolio started with \$1,000,000 and ended with \$1,328,581 with a Sharpe ratio of 1.09. The Sharpe ratio, is a measure of the risk-adjusted return of an investment or portfolio \cite{sharpe_ratio_origin}.

\begin{equation}
\text{Sharpe Ratio} = \frac{R_p - R_f}{\sigma_p}
\end{equation}
\begin{align*}
\text{where:} \\
R_p &= \text{return of portfolio} \\
R_f &= \text{risk-free rate} \\
\sigma_p &= \text{standard deviation of the portfolio's excess return}
\end{align*}

\begin{figure}[H]
      \centering
      \includegraphics[width=1\textwidth]{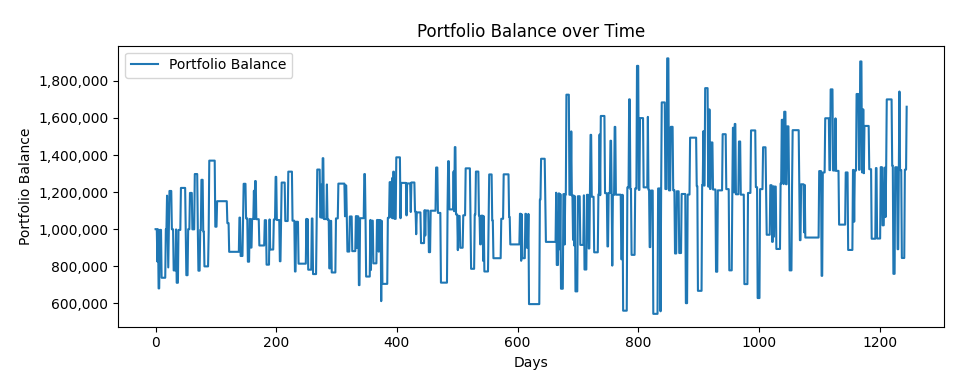}
      \caption{Portfolio balance over time.}
      \label{fig:sample-image}
\end{figure}

\begin{figure}[H]
      \centering
      \includegraphics[width=1\textwidth]{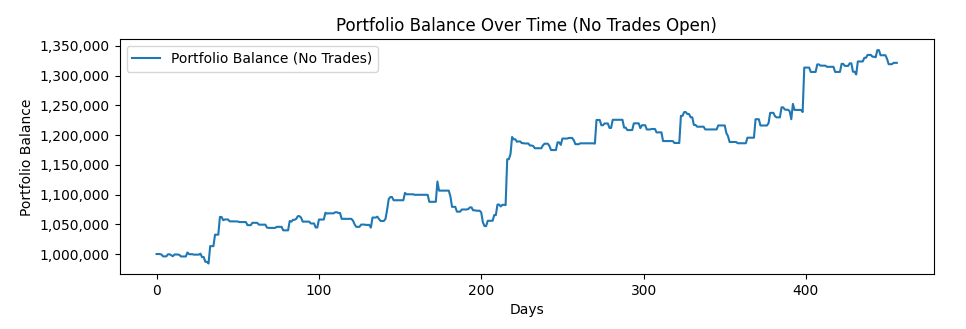}
      \caption{Portfolio balance over time with no trades open.}
      \label{fig:sample-image}
\end{figure}

\begin{figure}[H]
      \centering
      \includegraphics[width=1\textwidth]{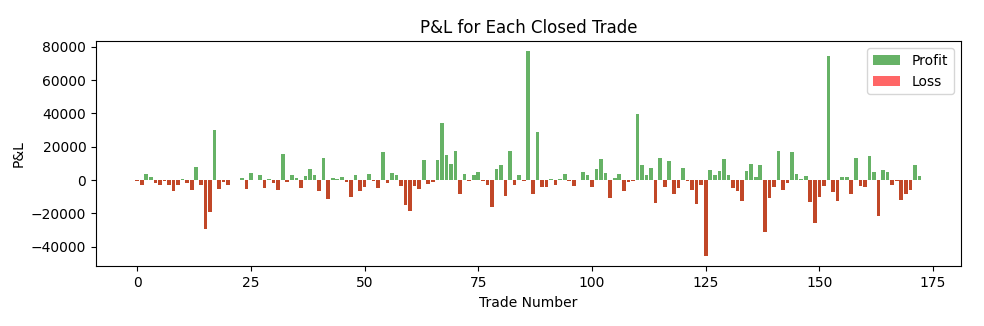}
      \caption{P\&L for each trade position.}
      \label{fig:sample-image}
\end{figure}

\subsection{Optimisation}
Utilising the low-latency programming repository created earlier, the pairs trading strategy was optimised to improve latency. This version incorporates optimisation at the CPU level by integrating Single Instruction, Multiple Data (SIMD) instructions, specifically using Advanced Vector Extensions 2 (AVX2) instructions. These instructions take advantage of the parallel processing power of contemporary CPUs, allowing simultaneous execution of multiple operations. A critical area of optimisation within the algorithm lies in the computation of the sum and sum of squares of the spread. In contrast to the traditional sequential approach, the optimised algorithm deploys AVX2 instructions to perform these computations in parallel. This parallelism significantly reduces the total number of iterations within the loop, thereby substantially accelerating computational speed. This enhancement has the caveat of being applicable only for window sizes that are multiples of four, a constraint that aligns well with the 256-bit wide vector operations performed by AVX2 instructions. This optimisation automatically includes loop unrolling within it. \\

\begin{lstlisting}[style=customc++, language=C++, caption={Calculations without SIMD and loop unrolling}]
    double mean = calc_mean(spread);
	double stddev = calc_stddev(spread, mean);
	double current_spread = stock1_prices[i] - stock2_prices[i];
	double z_score = (current_spread - mean) / stddev;  
\end{lstlisting}

\begin{lstlisting}[style=customc++, language=C++, caption={Calculations with SIMD and loop unrolling}]
	__m256d sum_vec = _mm256_setzero_pd();
	__m256d sq_sum_vec = _mm256_setzero_pd();

	for(size_t j = 0; j < N; j += 4) {
  	__m256d spread_vec = _mm256_loadu_pd(&spread[j]);
  	sum_vec = _mm256_add_pd(sum_vec, spread_vec);
  	sq_sum_vec = _mm256_add_pd(sq_sum_vec, _mm256_mul_pd(spread_vec, spread_vec));
	}

	__m256d temp1 = _mm256_hadd_pd(sum_vec, sum_vec);
	__m256d sum_vec_total = _mm256_add_pd(temp1, _mm256_permute2f128_pd(temp1, temp1, 0x1));

	__m256d temp2 = _mm256_hadd_pd(sq_sum_vec, sq_sum_vec);
	__m256d sq_sum_vec_total = _mm256_add_pd(temp2, _mm256_permute2f128_pd(temp2, temp2, 0x1));

	double sum = _mm_cvtsd_f64(_mm256_castpd256_pd128(sum_vec_total));
	double sq_sum = _mm_cvtsd_f64(_mm256_castpd256_pd128(sq_sum_vec_total));

    double mean = sum / N;
	double stddev = std::sqrt(sq_sum / N - mean * mean);

	double current_spread = stock1_prices[i] - stock2_prices[i];
	double z_score = (current_spread - mean) / stddev;
\end{lstlisting}

The algorithm further optimises through the incorporation of a fixed-size array to store the spread values. A rolling index strategy is implemented to keep track of and replace the oldest spread value in the array with the newest one. This approach negates the need for dynamic memory allocation, which would otherwise contribute to additional computational overhead. Consequently, this contributes to further enhancing performance. Finally the fourth optimisation was inlining both the functions for calculating the mean and standard deviation.

\subsection{Results}

Each optimisation used to improve the pairs trading strategy was tested in isolation to evaluate the individual performance of the design pattern before combining them together. The pairs trading strategy took 519772 nanoseconds without any optimisations. The individiual and combined optimisation can be see in Table 3 and Figure 18. A detailed evaluation and cache analysis can be seen in Section 6.2. 

\begin{table}[htbp]
\centering
\begin{tabular}{lccc}
\toprule
\textbf{Optimisation Technique} & \textbf{Latency (ns)} & \textbf{Improvement (\%)} \\
\midrule
Inlining & 406709  & 21.41 \\
SIMD and loop unrolling & 355618 & 31.28 \\
Fixed array & 266146 & 48.58 \\
Combined & 65580 & 87.33 \\
\bottomrule
\end{tabular}
\caption{Optimisation techniques and their performance improvements}
\label{tab:optimisations}
\end{table}

\begin{figure}[H]
      \centering
      \includegraphics[width=1\textwidth]{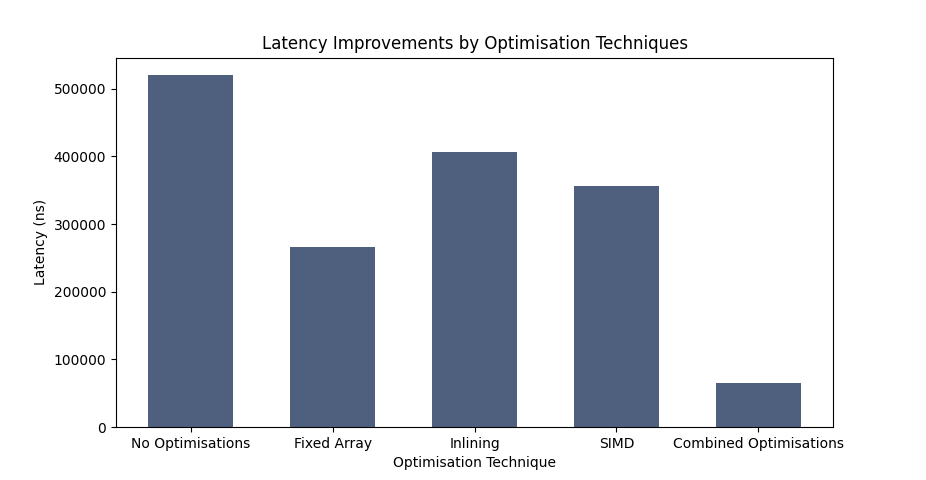}
      \caption{Latency improvements by optimisation techniques.}
      \label{fig:sample-image}
\end{figure}

\section{The Disruptor}
The third part of this work aimed to provide a tangible software library by building the LMAX Distruptor in C++. The library can be accessed from the same repository as the design patterns. The C++ Distruptor pattern was then tested against a regular queue structure to test improvement in latency and speed. The structure was also tested for cache analysis. 

\subsection{Core concepts}
\subsection*{Producer}
The producer in the Disruptor pattern is responsible for producing and publishing events \cite{13}. In its simplest form, the producer claims a slot in the ring buffer, writes data into it, and then makes it available to consumers by updating the sequence number. The primary advantage of this design is that, in many cases, only one producer is writing to a particular slot in the buffer at any given time. This ensures minimal contention and overhead, as atomic operations or locks are often unnecessary, maximizing the throughput.

\subsection*{Ring buffer}
The Ring Buffer is a circular data structure that holds the events \cite{13}. It is pre-allocated to a fixed size, ensuring that memory operations are both predictable and minimized. The pre-allocation helps to eliminate the need for garbage collection or memory allocation during runtime, contributing to the Disruptor's high performance. Events in the buffer are referenced by a sequence number, which always increments, ensuring each event has a unique slot in the buffer.

\begin{figure}[H]
      \centering
      \includegraphics[width=0.7\textwidth]{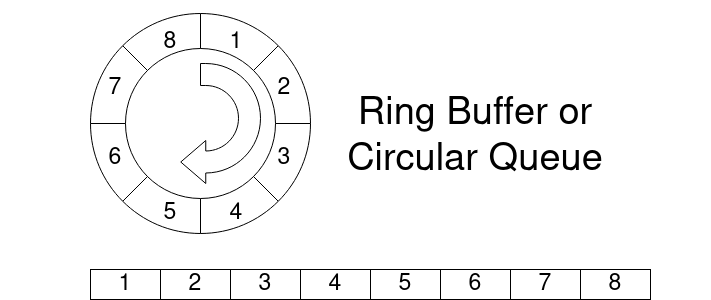}
      \caption{Ring buffer.}
      \label{fig:sample-image}
\end{figure}

\subsection*{Sequencer}
The sequencer is the mechanism by which producers and consumers can claim, track, and publish sequence numbers \cite{13}. It is effectively the synchronization point between producers and consumers. The sequence numbers determine which slot in the ring buffer is available for writing or has been written and is available for reading. The Sequencer can be viewed as the main controller that manages the buffer's access, ensuring that writes and reads are coherent and orderly.

\subsection*{Event processor}
The consumer, or often termed as the Event Processor, is responsible for consuming events from the ring buffer \cite{13}. The power of the Disruptor pattern lies in its ability to have multiple consumers work in parallel, or in sequence, without the need for locks. Each consumer tracks which events it has processed by sequence number. Some consumers can be set up to handle specific events, while others can process events only after certain other events have been handled, creating a dependency graph.

\subsection*{Sequence barrier}
The Sequence Barrier is a coordination mechanism used primarily by consumers to determine which sequences (events) they can safely read \cite{13}. By checking the sequence numbers provided by the Sequencer, the Sequence Barrier can tell a consumer if it's safe to read a particular event or if it needs to wait. If a consumer needs to wait, the Sequence Barrier will utilise the Wait Strategy to determine how to wait (busy spin, yield, sleep, etc.).

\subsection*{Event}
The event is a unit of data or object passed from the producer to the consumer \cite{13}. The event can be anything defined by the user. A HFT example of an event can be an order object which needs to be sent to the Order Management System (OMS). For simplicity sake a simple object containing a string was used but if the user wants to use the library they would probably like to change the object to their needs.

\subsection*{Wait Strategy}
This component decides how consumers should wait when there's no available event to be processed \cite{13}. There are various strategies for waiting, ranging from busy-spin waiting (where the consumer continually checks for availability) to more relaxed strategies like yielding or sleeping. The choice of strategy is crucial as it determines the trade-off between latency and CPU usage. For instance, a busy-spin strategy will provide low latency at the cost of high CPU usage, while a sleeping strategy will have higher latency but much lower CPU usage. \\
\\
In essence, the Disruptor is an orchestrated system where producers write events, and consumers process them with maximum efficiency and minimal contention. By removing the need for locks and taking full advantage of modern CPU caching and parallelism, the Disruptor achieves high performance that is often orders of magnitude faster than traditional queuing methods.

\subsection{Implementation}
\begin{figure}[H]
      \centering
      \includegraphics[width=0.8\textwidth]{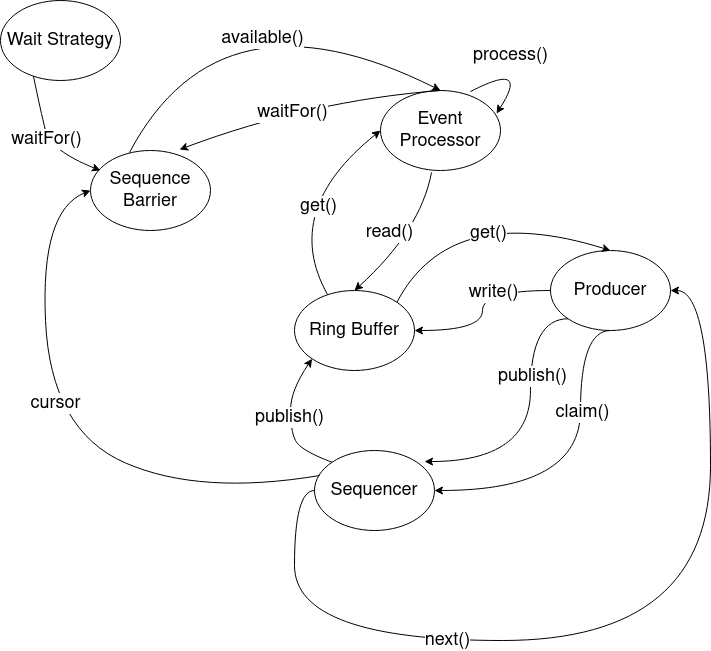}
      \caption{Flowchart of the Disruptor pattern implemented in C++ .}
      \label{fig:sample-image}
\end{figure}

\subsection{Performance comparison}
Two tests were run using Google Benchmark to assess the performance of inter-thread communication mechanisms between the Distruptor pattern and a simple queue.\\
\\
\texttt{BM\_Disruptor}: In this test, the Disruptor Pattern was leveraged. Key components of this pattern include a \texttt{RingBuffer} and a \texttt{Sequencer}. These shared components were instantiated, and a producer was set up to send \texttt{NUM\_EVENTS} to a consumer. The consumer was set up on a separate thread, and it processed these events. The producer pushed data into the \texttt{RingBuffer}, from where the consumer thread processed it in order.

\begin{lstlisting}[style=customc++, language=C++, caption={Disruptor Google Benchmark}]
static void BM_Disruptor(benchmark::State& state) {
    for (auto _ : state) {
        // Create shared instances of the RingBuffer and Sequencer
        auto ringBuffer = std::make_shared<RingBuffer>(NUM_EVENTS);
        auto sequencer = std::make_shared<Sequencer>();

        // Create the producer and consumer
        Producer producer(ringBuffer, sequencer);
        EventProcessor consumer(ringBuffer, sequencer);
        std::thread consumerThread([&consumer]() { consumer.run(); });

        for (int i = 0; i < NUM_EVENTS; ++i) {
            producer.onData("Event " + std::to_string(i));
        }

        // Stop the consumer and wait for it to finish
        consumer.stop();
        consumerThread.join();
    }
}
BENCHMARK(BM_Disruptor);
\end{lstlisting}

%BM\_SimpleQueue: Here, a standard queue from the C++ Standard Library was utilised for inter-thread communication. A mutex and a condition variable were employed to ensure thread-safe operations and synchronization. The producer pushed NUM\_EVENTS into this queue, while the consumer, running on a separate thread, waited for the condition variable to be notified, after which it consumed the events.

\texttt{BM\_SimpleQueue}: Here, a standard queue from the C++ Standard Library was utilised for inter-thread communication. A mutex and a condition variable were employed to ensure thread-safe operations and synchronization. The producer pushed \texttt{NUM\_EVENTS} into this queue, while the consumer, running on a separate thread, waited for the condition variable to be notified, after which it consumed the events.

\begin{lstlisting}[style=customc++, language=C++, caption={Simple Queue Google Benchmark}]
static void BM_SimpleQueue(benchmark::State& state) {
    for (auto _ : state) {
        std::queue<std::string> queue; 
        std::mutex mutex;
        std::condition_variable condVar;
        bool done = false;

        std::thread consumerThread([&] {
            while (!done) {
                std::unique_lock<std::mutex> lock(mutex);
                while (queue.empty() && !done) {
                    condVar.wait(lock);
                }
                if (!queue.empty()) {
                    std::string event = queue.front(); 
                    queue.pop();
                }
            }
        });

        for (int i = 0; i < NUM_EVENTS; ++i) {
            std::string event = "Event " + std::to_string(i); 
            {
                std::lock_guard<std::mutex> lock(mutex);
                queue.push(event);
            }
            condVar.notify_one();
        }

        done = true;
        condVar.notify_one();
        consumerThread.join();
    }
}
BENCHMARK(BM_SimpleQueue);
\end{lstlisting}

\subsection{Results}

\texttt{BM\_Disruptor} (using the Disruptor Pattern) was considerably faster, on average it was \(38\%\) faster compared to the \texttt{BM\_SimpleQueue}. A detailed analysis, statistical testing, and cache analysis of the results can be seen in the evaluation in Section~6.3.

\begin{table}[htbp]
\centering
\begin{tabular}{lccc}
\toprule
\textbf{Events} & \textbf{Queue Speed (ns)} & \textbf{Disruptor Speed (ns)} & \textbf{Speed up (\%)} \\
\midrule
10 & 20646 & 18182 & 11.9 \\
100 & 99458 & 64686 & 35.0\\
1000 & 881092 & 451251 & 48.8\\
10000 & 9735102 & 4361096 & 55.2\\
100000 & 90088609 & 52562872 & 41.7\\
1000000 & 884871405 & 543171556 & 38.7\\
\bottomrule
\end{tabular}
\caption{Performance comparison between Queue and Disruptor.}
\label{tab:performance_comparison}
\end{table}

\noindent The speed increase observed with the Disruptor Pattern can be attributed to its design principles:

\begin{enumerate}
    \item \textbf{Avoiding Lock Contention:} The Disruptor Pattern is designed to avoid lock contention which is inherent in traditional producer-consumer models using mutexes and condition variables. The \texttt{SimpleQueue} test incurred overhead each time the mutex was locked and unlocked, and the condition variable was used for synchronization. The Disruptor Pattern avoids this overhead.
    \item \textbf{Memory Access Patterns:} The RingBuffer provides a data structure that utilises memory more efficiently. Its circular nature ensures cache lines are better utilised, leading to fewer cache misses and better performance.
    \item \textbf{Predictable Flow:} Unlike standard queues where memory allocation and deallocation can be unpredictable, the RingBuffer in the Disruptor Pattern provides a predictable flow. The same memory space is reused for events, which reduces the overhead of memory operations.
    \item \textbf{Decoupling of Producer and Consumer:} In the Disruptor model, the producer and consumer are decoupled more effectively than in simple queue-based models, enabling better parallelism and reducing potential bottlenecks.
\end{enumerate}

\noindent In summary, the Disruptor Pattern takes advantage of modern CPU architecture, optimising cache usage, and reducing lock contention, leading to its superior performance over the traditional producer-consumer setup using standard queues.

\section{Evaluation}

\subsection{Repository evaluation}

Participants were invited to explore the Low-Latency Programming Repository by diving into the introduction on the homepage, followed by a deep dive into each technique mentioned in the archive's articles/section. A total of 17 individuals, with varying ages, participated from four universities, Imperial College London, University College London (UCL), King's College London, and University of Oxford. 11 out of the 17 participants were computer science students whilst the remaining 6 were a mixture of mathematics, physics and engineering students. Participants then ranked the archive based on these distinctive metrics. Comprehensiveness: The range of topics discussed. Clarity: The lucidity of the explanations provided.

\subsubsection*{Comprehensiveness}
This metric was derived from the participants' views on the variety of topics the archive showcased. The average score suggests that users appreciated the wide range of topics present in the archive. However, some participants indicated a desire for more in-depth discussions on data handling and concurrency topics as they had the least amount of optimisations techniques.

\subsubsection*{Clarity}
Clarity evaluates how clearly and understandably topics are presented in the archive. This encompasses if the ideas were highlighted concisely without being ambiguous. The commendable average score shows that users found the archive to be clear and easy to follow. One evaluator noted that each benchmark code included a simple to understand example, making the ``content easy to digest and practical." However, a critique raised was about the lack of comments used in some of the examples. More descriptive comments can be add to make sure the benchmark is clear in its objective.

\begin{table}[htbp]
\centering
\begin{tabular}{lccc}
\toprule
\textbf{Categories} & \textbf{Average Score} & \textbf{Lowest Score} & \textbf{Highest Score} \\
\midrule
Comprehensiveness & 8.3 & 6.5 & 9.1  \\
Clarity & 8.7 & 7.8 & 8.9  \\
\bottomrule
\end{tabular}
\caption{User feedback scores}
\label{tab:optimisations}
\end{table}

\noindent In summary, while the archive was widely praised for its extensive coverage and clear explanations, there are areas for improvement.

\subsubsection*{Reproducibility of repository}
Ensuring the reproducibility of research is essential for transparency and the advancement of knowledge. This section highlights the steps taken in the testing process. The provided repository contains the code, benchmark configurations, and data necessary for other researchers to validate reported findings and build upon this work. The repository is organized in a clear and intuitive structure that enables easy navigation and understanding of the testing process. Each technique's benchmarking code is housed within its own dedicated directory. Each code file includes inline comments explaining the purpose of specific functions, benchmarks, and measurements. Detailed comments clarify the intended behavior of the code and how different parts of the technique are implemented. This documentation serves as a guide for researchers aiming to understand, modify, or extend our work. The repository includes configuration files for the benchmarking tool used (e.g., Google Benchmark). These files specify the parameters, test cases, and iterations for each benchmark scenario. Researchers can use these configurations to precisely replicate our benchmarking experiments. However if the reader has not installed Google Benchmark then they must so before running the benchmarks in the repository. Google Benchmark installation can be found at \href{https://github.com/google/benchmark}{\textcolor{black}{\texttt{https://github.com/google\\/benchmark}}}. After Google Benchmark has been installed and the benchmark folder if in the same directory as the mybenchmark.cc file, the user can compile the mybenchmark.cc file using the makefile provided in the repository. The makefile uses g++ as the compiler so the user must change this accordingly to what they are using on their system. By running the executable file, mybenchmark, the user is able to observe the results of the benchmark directly from their terminal. While significant steps have been taken to enhance reproducibility, it's important to acknowledge potential challenges and limitations. Variability in hardware configurations, compiler versions, and operating systems can influence results. However, we have strived to document these variables and provide guidelines for minimizing their impact.

\subsection{Trading evaluation}

\subsubsection*{Speed analysis}

The programming techniques combined improved the latency of the trading algorithm by 87.38\%. The total speed of the non-optimised pairs trading took 517559.10 nanoseconds on average whilst the optimised took 65588.30 nanoseconds on average. Testing was done with 10 times running both versions. The mean and standard deviation, nearest to the whole nanosecond, can be seen in Table 6. The smaller standard deviation in the optimised version indicates that the latency values are more tightly clustered around the mean, which can contribute to a more predictable and consistent performance.

\begin{table}[htbp]
\centering
\begin{tabular}{lccc}
\toprule
\textbf{Metric} & \textbf{Non-Optimised} & \textbf{Optimised} \\
\midrule
Mean Latency (ns) & 517559 & 65588  \\
Standard Deviation (ns) & 4233 & 400  \\
\bottomrule
\end{tabular}
\caption{Mean and standard deviation of trading optimised and non-optimised.}
\label{tab:optimisations}
\end{table}

The speed ranges for the 10 tested benchmarks for optimised and non-optimised code can be seen in as box plots in Figure 21. The spread is greater in the non-optimised version due to an anomaly which could potentially imply that the optimisations not only improved speed but also uniformed performance. However further testing would be required to draw a conclusion.

\begin{figure}[H]
      \centering
      \includegraphics[width=0.6\textwidth]{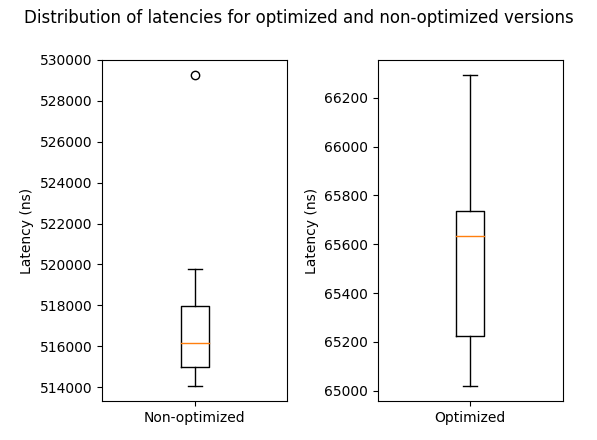}
      \caption{Box plots of latency for optimised and non-optimised trading algorithms.}
      \label{fig:sample-image}
\end{figure}

Each optimisation used to improve the pairs trading strategy was tested in isolation to evaluate the individual performance of the design pattern before combining them together. Each benchmark was tested 10 times. The averages can be seen in Table 7.

\begin{table}[htbp]
\centering
\begin{tabular}{lccc}
\toprule
\textbf{Optimisation Technique} & \textbf{Latency (ns)} & \textbf{Baseline (ns)} & \textbf{Improvement (\%)} \\
\midrule
Inlining & 406709 & 517559 & 21.41 \\
SIMD and loop unrolling & 355618 & 517559 & 31.28 \\
Buffer (fixed array) & 266146 & 517559 & 48.58 \\
Combined optimisation & 65580 & 517559 & 87.33 \\
\bottomrule
\end{tabular}
\caption{Trading sub-optimisation techniques and their performance improvements}
\label{tab:optimisations}
\end{table}

\begin{figure}[H]
      \centering
      \includegraphics[width=0.8\textwidth]{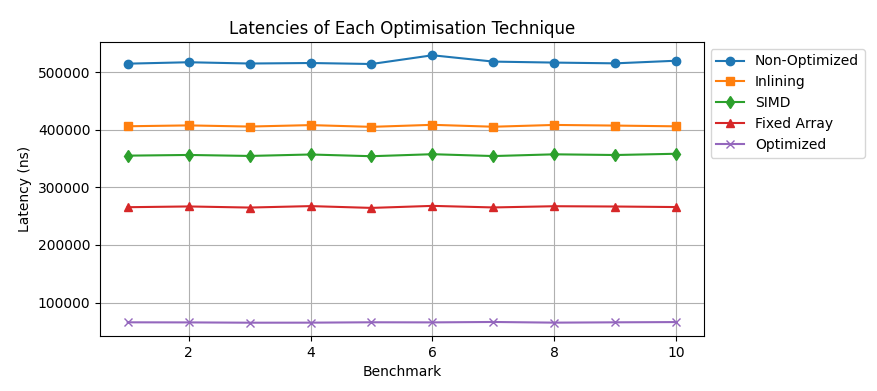}
      \caption{Latencies of each optimisation techinique.}
      \label{fig:sample-image}
\end{figure}

Furthermore, Figure 23 presents violin plots for each optimisation technique, highlighting the distribution of speeds associated with each method. The principal conclusion that can be drawn is that the distribution of speeds for the non-optimised code is skewed in comparison to the distributions for the optimised benchmarks. Specifically, the shape of the violin plot for the non-optimised code indicates a greater concentration of data points at the lower end of the range. The increased density in this region suggests that these lower speeds are more prevalent or typical for this particular dataset. This skewness could potentially be attributed to anomalies in the data.

\begin{figure}[H]
      \centering
      \includegraphics[width=1\textwidth]{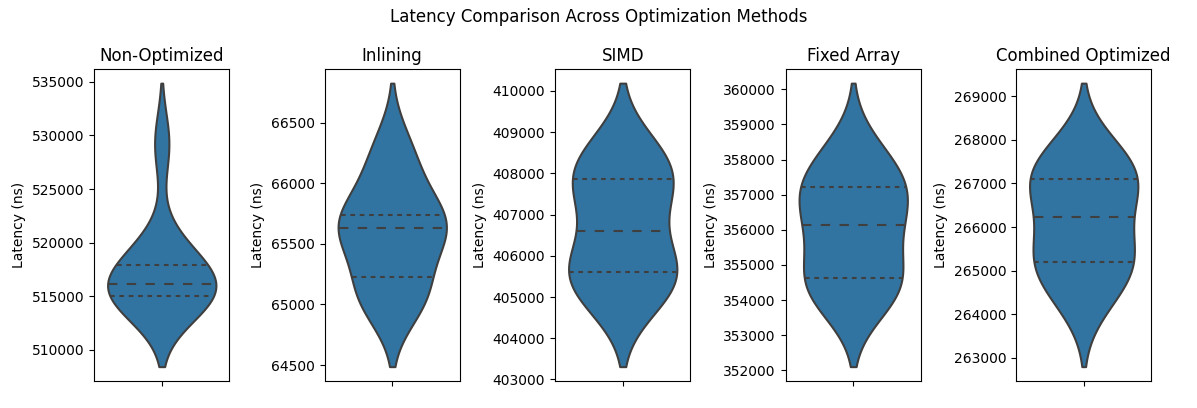}
      \caption{Violin plots of latency comparison across optimisation methods.}
      \label{fig:sample-image}
\end{figure}

\subsubsection*{Cache analysis}
Among the optimisation techniques evaluated, buffer (fixed array), inlining, and SIMD, all demonstrated an increase in the total number of instructions compared to the original version. Interestingly, despite having higher instruction counts, their performances were not uniform, indicating the importance of other factors like cache utilisation and execution efficiency.

\begin{itemize}
    \item \textbf{Buffer Data Structure:} Even with its higher number of instructions, this technique significantly outperformed the original. This implies a more efficient execution of instructions, possibly aided by optimised memory access patterns.
    \item \textbf{Inlining:} Again inlining too has a higher number of instructions however it only slightly outperforms the original.
    \item \textbf{SIMD:} Like the other techniques, SIMD also had a higher number of instructions but delivered better performance than the original. However, its efficiency seems to be dependent on other factors like parallelism and data-level operations.
\end{itemize}

\begin{table}[htbp]
\centering
\begin{tabular}{lc}
\toprule
\textbf{Optimisation Technique} & \textbf{Number of Instructions (Billion)} \\
\midrule
Without optimisation & 6.01 \\
Combined optimisation & 3.27 \\
Buffer (fixed array) & 8.27 \\
Inlining & 9.07 \\
SIMD and loop unrolling & 8.35 \\
\bottomrule
\end{tabular}
\caption{Number of instructions for each optimisation techniques}
\label{tab:num_instructions}
\end{table}

The cache miss rate offers crucial insights into the performance of each technique:

\begin{itemize}
    \item \textbf{Buffer Data Structure:} Though it had a higher cache miss rate compared to the original, it was still lower than that of the fully optimised version. This suggests a balanced approach to memory access, contributing to its overall effectiveness.
    \item \textbf{Inlining:} This technique had a cache miss rate similar to that of the Buffer Data Structure, but its overall performance was inferior, indicating less efficiency in memory access or data retrieval.
    \item \textbf{SIMD:} Intriguingly, SIMD had the same cache miss rate as the original, indicating that it doesn't significantly affect cache performance when used alone. 
\end{itemize}

\begin{table}[htbp]
\centering
\begin{tabular}{lc}
\toprule
\textbf{Optimisation Technique} & \textbf{Cache Misses (\% of all cache refs)} \\
\midrule
Without optimisation & 16.001 \\
Combined optimisation & 33.879 \\
Buffer (fixed array) & 19.089 \\
Inlining & 19.151 \\
SIMD and loop unrolling & 16.829 \\
\bottomrule
\end{tabular}
\caption{Cache misses for each optimisation techniques}
\label{tab:cache_misses}
\end{table}

\subsubsection*{Statistical significance}
The t-test serves as a statistical hypothesis test designed to compare the means of two distinct groups for the purpose of determining the statistical significance of any observed differences \cite{statistics_principles}. This test is conducted under the premise of the null hypothesis, which asserts that no significant difference exists between the groups in question. The outcome of the t-test is expressed through a t-statistic, a standardized indicator quantifying the scale of the observed difference in relation to the variability inherent in the dataset \cite{t_test_handbook}. Additionally, a p-value is calculated to represent the likelihood of encountering an outcome as extreme as the observed result under the assumption that the null hypothesis holds true. A lower p-value provides more compelling evidence against the null hypothesis, thereby suggesting a significant difference between the groups under scrutiny. In the specific context of evaluating latency improvements in optimised system versions, a paired t-test was employed. This variant of the t-test is particularly suited for observations collected in pairs, such as pre- and post-optimisation measurements on an identical unit of observation. 

The 10 latency measurements were used for both the non-optimised and optimised versions of the trading strategy. The t-statistics and p-values are presented in Figure 24. The data reveals that the t-statistics were notably large, indicating a significant discrepancy between the non-optimised and optimised conditions. Concurrently, the p-values for each group were extremely low, thereby implying that the probability of the observed latency improvements being a product of random variation is minimal. Given these findings, there is strong statistical support for the assertion that the implemented optimisation strategies have effectively reduced system latency. Consequently, it may be conclusively stated that the optimisation efforts have yielded a significant enhancement in system performance. However to improve accuracy it would be advised to use a greater number of data points as 10 could potentially be limited.

\begin{figure}[H]
      \centering
      \includegraphics[width=0.7\textwidth]{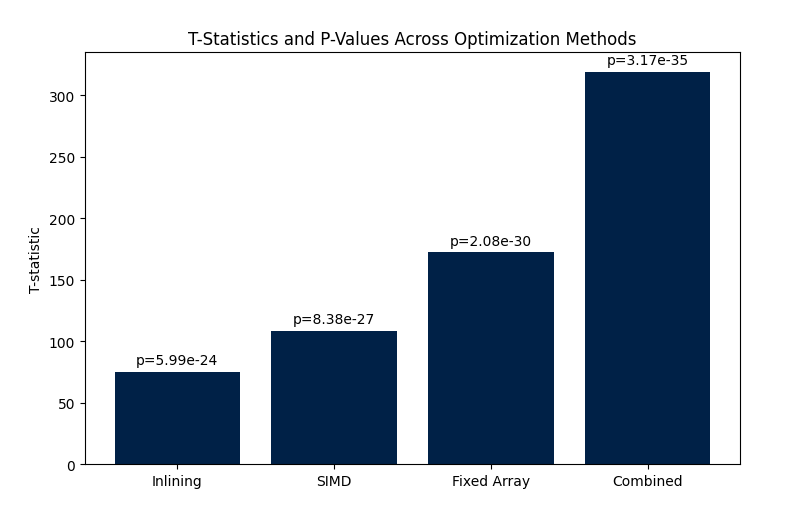}
      \caption{T-statistics and p-values across optimisation methods.}
      \label{fig:sample-image}
\end{figure}

\subsubsection*{Profitability analysis}

The main reasoning behind reducing latency is the aim of increasing trading profit. Reduced latency provides a decisive advantage by allowing opportunities, such as fleeting arbitrage gaps between exchanges or adjusting bid-ask prices in market-making, faster than competitors.

In the study conducted by Baron \cite{baron2019}, empirical evidence indicates that HFT firms realize a disproportionately high level of revenue characterized by elevated Sharpe ratios. This achievement is attributed to their engagement in trades with a broad spectrum of market participants over concise time intervals. Moreover, the firms with the fastest trading capabilities consistently exhibit superior performance. The enhanced performance manifests through two distinct channels: the short-lived information pathway and the risk management pathway. Furthermore, rapidity proves beneficial across a range of strategies, notably in market making and cross-market arbitrage \cite{baron2019}. Speed can be categorized into two main sections, relative speed and absolute speed. In this work, ``relative speed" denotes speed when benchmarked against the quickest HFT firm. Thus, any enhancement in relative speed is in comparison to other HFT entities. In contrast, ``absolute speed" stands independent and is not assessed in relation to any other benchmark.

Relative speed increase can be pivotal for firms aiming to maximize their profits. This is evidenced by research that draws a connection between the latency in HFT and trading profitability. Baron, elucidate that the fastest HFT firms often report superior performance when compared to their slower counterparts \cite{baron2019}. Interestingly, they employ the Herfindahl index to evaluate market concentration, with a higher number signifying a market dominated by a few major players and a lower one indicating a fragmented market. Research conducted by Van Ness et al. in 2005 using NASDAQ data observed a significant indicator: despite the growth and influx of participants in the HFT sector, a consistent profitability was maintained by a limited number of firms with superior technological advantages \cite{vanNess2005}. This phenomenon underscores the importance of speed in this trading segment.

As the pairs trading strategy in Section 4 was dealing with absolute speeds the focus will be on how absolute speeds increase profitability. Delving into the mechanics of latency arbitrage strategies can provide a clearer picture. These strategies capitalize on speed advantages to harness price discrepancies between similar financial assets across distinct markets. Wah \cite{wah2013} provides a mathematical blueprint for understanding such opportunities, essentially revolving around crossed market scenarios, where bids in one market surpass asks in another, and the timeframes for these discrepancies allow potential exploitation. The main idea for Latency Arbitrage Opportunity (LAO) explained by Wah is as follows \cite{wah2013}:

\begin{itemize}
    \item \textbf{Bid (B):} Represents the price at which a market participant is willing to purchase a security.
    \item \textbf{Ask (A):} Denotes the price at which a market participant is prepared to sell a security. The differential between the bid and ask prices is conventionally termed as the 'spread'.
\end{itemize}

An LAO is discerned when the following conditions concurrently manifest:
\begin{enumerate}
    \item \textbf{Crossed Market Condition:} This arises when the bid price from Market 1 (BM1) surpasses the ask price from Market 2 (AM2). Such a scenario begets an arbitrage avenue, allowing a trader to purchase the asset in Market 2 at its lower ask price and contemporaneously offload it in Market 1 at the superior bid price, thus capturing the price differential as profit.
    \item \textbf{NBBO Comparison:} The Securities Exchange Commission (SEC) has instituted the NBBO regulation to ensure that investors are accorded the most favorable ask price when procuring securities and the optimal bid price during divestiture. For a latency arbitrage opportunity to be deemed viable, the bid price in Market 1 should be on par with or exceed the national best bid (NBB), and concurrently, the ask price in Market 2 must be less than or equivalent to the national best offer (NBO). This stipulation ensures that the price dynamics from Markets M1 and M2 are at least as propitious as the best extant prices disseminated across all national exchanges.
    \item \textbf{Time Viability Condition:} To ensure the practical exploitability of the observed price discrepancy, there must be a positive time interval from the inception to the cessation of this differential. While these windows of opportunity in HFT are evanescent, often measured in mere milliseconds, the unparalleled trading velocities of HFT systems permit them to capitalize on such transient disparities.
\end{enumerate}

\noindent A 2011 study by Ende et al. provides an intriguing perspective on this topic \cite{ende}. Rather than the customary focus on the direct impact of reduced latency on increased earnings, their research delves into the relationship between trading speed and susceptibility to adverse selection or disadvantageous shifts in the order book. By examining an active trading strategy across varied latencies for a specific stock on the German stock exchange, the authors found, through regression analysis, that a 1\% increase in latency corresponds to a 0.9\% heightened likelihood of exposure to undesired order book alterations. 

\begin{figure}[H]
      \centering
      \includegraphics[width=0.5\textwidth]{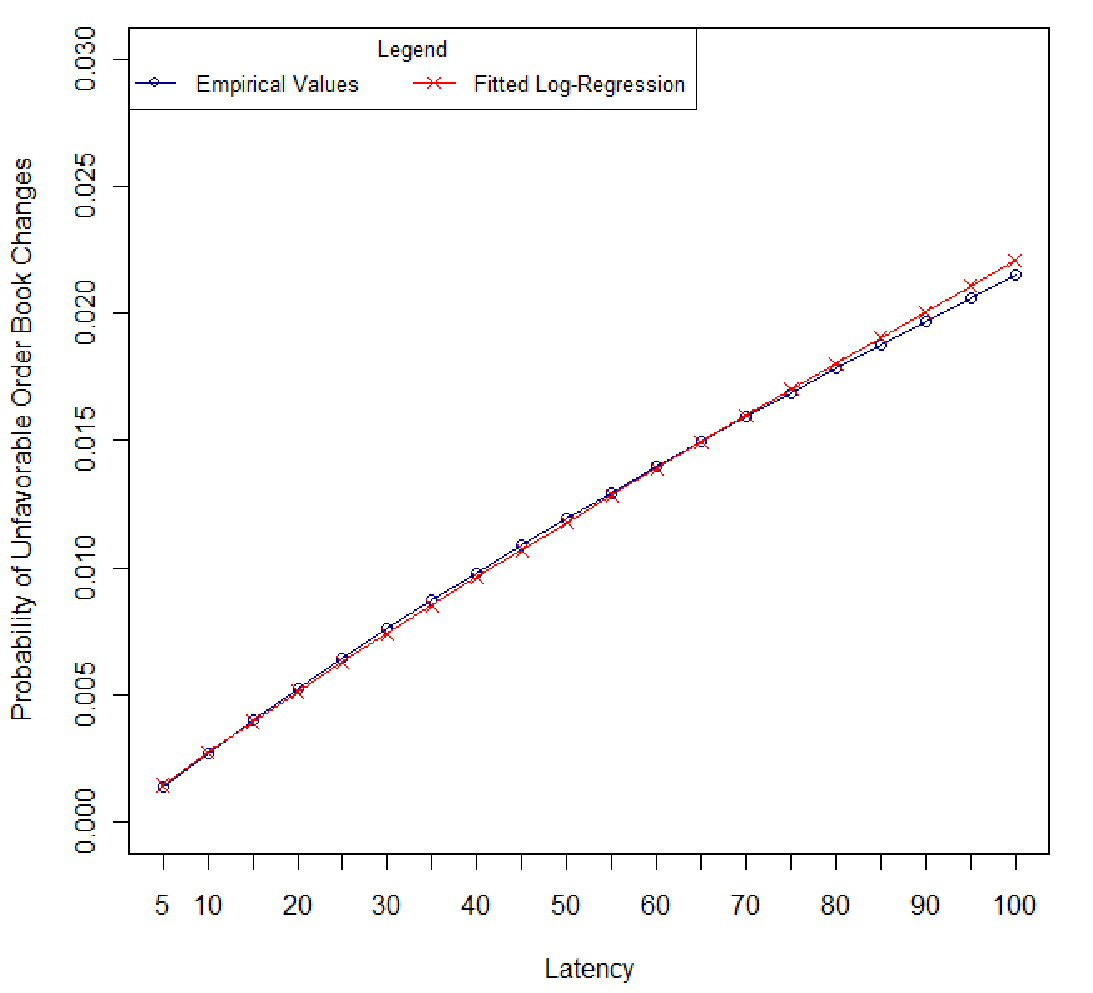}
      \caption{Probability of unfavourable order book changes against latency change \cite{ende}.}
      \label{fig:sample-image}
\end{figure}

\noindent As shown in Section 4.5, we can increase the execution speed of our trading algorithm by approximately 87.32\%. This would lead to a decrease of approximately 78.59\% in our exposure to unfavorable order book changes. The reset of the percentages for profitability can be seen in Table 10.

\begin{table}[H]
\centering
\begin{tabular}{lc}
\toprule
\textbf{Optimisation Technique} & \textbf{Exposure to unfavorable order book changes (\%)} \\
\midrule
Buffer (fixed array) & 43.72 \\
Inlining & 19.27 \\
SIMD and loop unrolling & 28.15 \\
Combined optimisation & 78.59 \\
\bottomrule
\end{tabular}
\caption{Exposure to unfavorable order book changes for each optimisation technique}
\label{tab:cache_misses}
\end{table}

\subsubsection*{Limitations}
The Benchmark conducted for the pairs trading backtesting had its limitations. Firstly backtesting speeds do not accurately portray speeds in production. As this was a backtesting code the prices were loaded into an vector before running the trading algorithm. However in a real world example this would not be the case. While backtesting can provide an accurate representation of past stock values, it does not account for the unpredictable nature and anomalies of live trading data. Live data streams can be affected by many factors such as network latency, unexpected market news, and irregular trading patterns. Therefore, while the algorithm might perform optimally in a controlled environment, its efficiency might vary when exposed to real-world, real-time conditions. One of the constraints introduced by the optimisation strategy tested in this study is the requirement for window sizes that are multiples of four. This constraint could limit the versatility of the algorithm, especially when dealing with diverse data sets or if specific non-multiple window sizes are preferred for certain trading strategies. Finally the optimisations which rely on AVX2 instructions, assume the presence of specific hardware capabilities. Thus, the performance benefits might not be realized if deployed on systems lacking the required hardware support.

\subsection{Distruptor evaluation}

To assess the performance advantages of the Disruptor pattern in comparison to a traditional queue, a series of benchmarks were conducted, targeting the speed of inter-thread communication, as well as cache analysis. Although these findings offer insights into the potential speed enhancements attributed to the Disruptor, it is crucial to discuss and contextualize certain nuances and facets of the performance evaluation.

\subsubsection*{Speed analysis}

As mentioned in Section 5.4 the Disruptor pattern was considerably faster, nearly two times faster  compared to the Simple Queue. To further investigate the results and how speed changes with varying number of events further tests were conducted. Each test was conducted 20 times with varying number of events. The results can be seen in Figure 26. An important conclusion can be drawn from the results. With low number of events to process the gap between the Disruptor and the Queue were closer. However as the number of events increased the gap between the two data structures become greater. At 10 events the average speed gap between the two structures was 2464 nanoseconds. At 1000000 events the average speed gap between the two structures was 341699849 nanoseconds.  This conclusion aligns strongly with the literature review from LMAX Group.

\begin{figure}[H]
      \centering
      \includegraphics[width=0.9\textwidth]{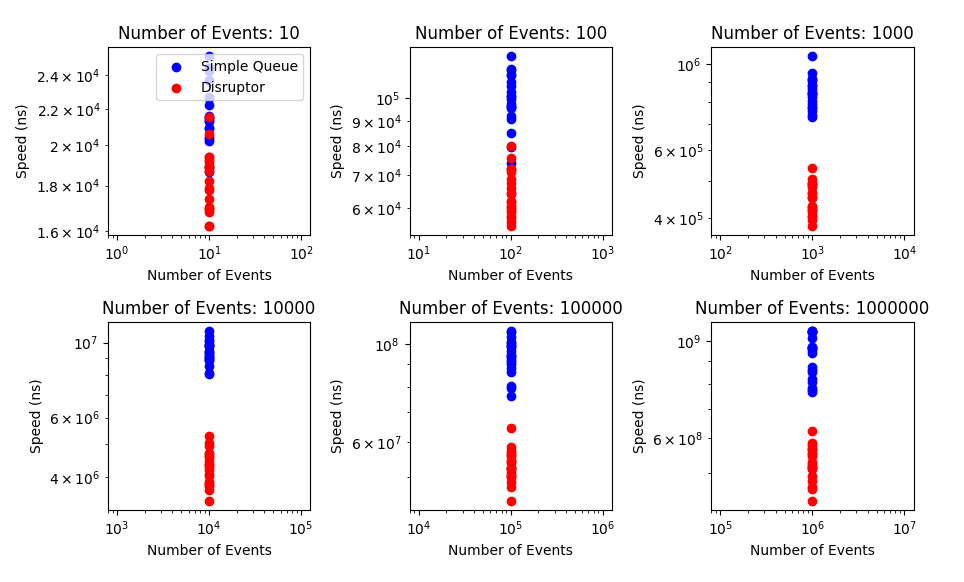}
      \caption{Speeds of Disruptor pattern and simple queue with varying number of events.}
      \label{fig:sample-image}
\end{figure}

\begin{figure}[H]
      \centering
      \includegraphics[width=0.9\textwidth]{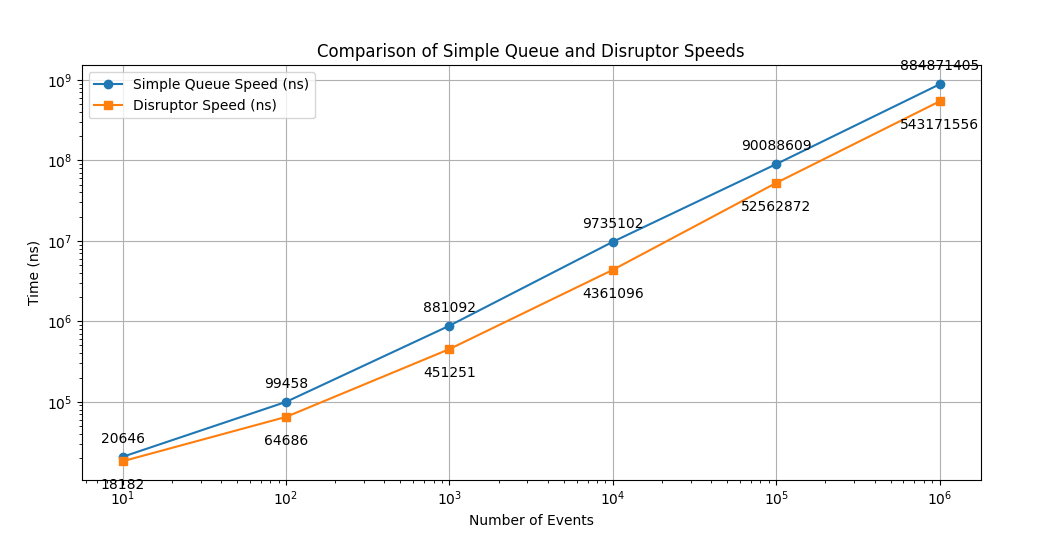}
      \caption{Comparison of simple queue and Disruptor speeds.}
      \label{fig:sample-image}
\end{figure}

\begin{figure}[H]
      \centering
      \includegraphics[width=0.9\textwidth]{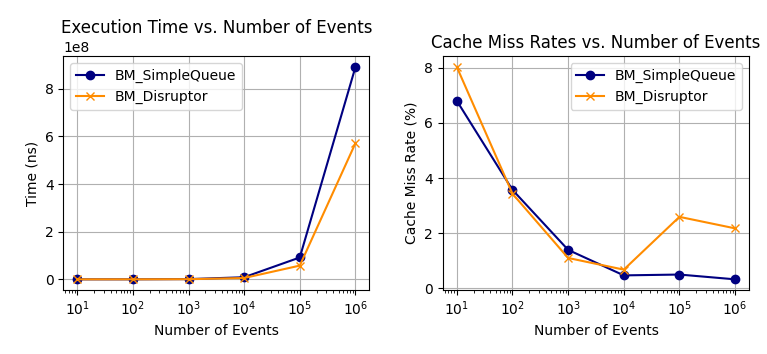}
      \includegraphics[width=0.5\textwidth]{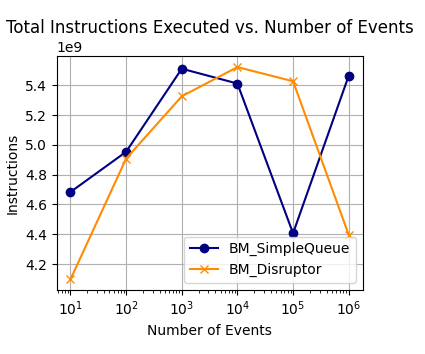}
      \caption{Cache analysis of Disruptor pattern.}
      \label{fig:sample-image}
\end{figure}

\subsubsection*{Cache analysis}
Cache efficiency, an essential factor in the data structure's performance, presents a more nuanced picture. At smaller event sizes---specifically \(10\) and \(100\) events---\texttt{BM\_Disruptor} exhibits a marginally higher cache-miss rate compared to \texttt{BM\_SimpleQueue}. However, this dynamic flips when examining larger event sizes ranging from \(1,000\) to \(1,000,000\). In these cases, \texttt{BM\_Disruptor} achieves lower cache-miss rates, implying enhanced cache efficiency. Overall, both data structures maintain relatively low cache-miss rates, but \texttt{BM\_Disruptor} shows better efficiency at scale.

In terms of instruction execution, \texttt{BM\_Disruptor} shows a propensity for operational efficiency. For example, when tested with \(10\) events, \texttt{BM\_Disruptor} executed approximately \(586,000\) fewer instructions than \texttt{B\_SimpleQueue}. This reduced instruction footprint can significantly contribute to better performance, especially when scaled to larger event sizes.

\subsubsection*{Statistical significance}

To test for statistical significance a total of \(20\) measurements with \texttt{NUM\_EVENTS} set to \(1000\) was tested. The mean and standard deviation for the testing can be seen in Table \(11\). The results from the statistical tests indicate a significant performance difference between the Simple Queue and the Disruptor when handling \(1000\) events, with the Disruptor showing markedly lower mean latency and standard deviation.

\begin{table}[htbp]
\centering
\begin{tabular}{lccc}
\toprule
\textbf{Metric} & \textbf{Simple Queue} & \textbf{Disruptor} \\
\midrule
Mean Latency (ns) & 931255 & 74908  \\
Standard Deviation (ns) & 453766 & 53600  \\
\bottomrule
\end{tabular}
\caption{Mean latency and standard deviation of Disruptor and queue.}
\label{tab:optimisations}
\end{table}

\noindent Furthermore, the t-statistic of 22.596 and an extremely low p-value of \(1.243 \times 10^{-23}\) strongly confirm that this performance difference is statistically significant, virtually eliminating the possibility that the observed differences are due to random variation. T-test results can be seen in Table 12. 

\begin{table}[htbp]
\centering
\begin{tabular}{lccc}
\toprule
\textbf{Test} & \textbf{t-statistic} & \textbf{p-value} \\
\midrule
T-Test (Queue v. Disruptor) & 22.596 & \(1.243 \times 10^{-23}\)  \\
\bottomrule
\end{tabular}
\caption{T-test results of Disruptor vs simple queue}
\label{tab:optimisations}
\end{table}

\subsubsection*{Limitations}

The benchmarks focused predominantly on the speed of inter-thread communication, a critical facet in concurrent programming paradigms. The inherent design and structure of the Disruptor pattern seem to provide significant advantages in this domain, as observed from the tests.

\noindent Nevertheless, performance evaluation encompasses more than just speed. In real-world applications, particularly in contexts like high-frequency trading systems or other latency-sensitive environments, several performance metrics are essential:

\begin{itemize}
\item \textbf{Memory Consumption}: Memory utilisation, capturing both peak and average consumption, holds significant importance. It is imperative for systems with constrained or vital memory resources to ensure that software components exhibit efficiency in memory consumption while maintaining speed.

\item \textbf{CPU Load}: Understanding the impact of software components on CPU load becomes essential, especially when considering scenarios with variable data sizes or heightened concurrent operations. This analysis ensures system robustness during peak loads without performance deterioration.
\end{itemize}

\noindent For an all-encompassing grasp of the Disruptor's performance and its relevance to particular scenarios, a more extensive evaluation of these factors is warranted.

The benchmark briefly addressed a fundamental component of the Disruptor pattern: the Wait Strategy. This component is pivotal to the functioning of the Disruptor as it stipulates the behavior of consumers when events are not available for processing. Consequently, the Wait Strategy exerts direct influence over latency and CPU usage. Though various wait strategies were acknowledged—ranging from aggressive busy-spin waits to more tempered, resource-saving methods—the specific strategy employed during the benchmark was yield wait. However, the differential impacts of the other listed strategies were not investigated. The chosen wait strategy can significantly determine the latency profile of the Disruptor. For instance, a busy-spin may provide minimal latency but at the cost of elevated CPU usage. On the other hand, a sleep strategy could result in marginally increased latency but might conserve CPU resources. A holistic comprehension of the Disruptor's performance capabilities necessitates a detailed comparison and analysis across diverse wait strategies. Without this, latency measurements might present a limited perspective, potentially influenced by the selection of a particular strategy.

\section{Conclusion and future directions}
\subsection*{Conclusion}
This work has aimed to bring transpacy to C++ design patterns used in the quantitative trading industry. This goal was achieved by the main three deliverables, the Low-Latency Programming Repository, the use of said repository in optimsing a statistical arbitrage pairs trading strategy, and finally developing LMAX Group's Disruptor pattern in C++ wiht the intension of being used in HFT Order Management Sytems. Each of these contributions were throughly evaluated. 

The repository's strategies showed positive results with Cache Warming and Constexpr showing the most significant gains at approximately 90\% each. Techniques like Loop Unrolling, Lock-Free Programming, and Short-circuiting also yielded strong results, while slowpath removal and signed vs unsigned comparisons were least impactful, improving speeds by around 12\%. Participants from four universities evaluated the Low-Latency Programming Repository on two key metrics: Comprehensiveness and Clarity, with average scores of 8.3 and 8.7 out of 10, respectively. While the repository was generally well-received for its wide range of topics and clear explanations, some participants expressed a desire for more in-depth content on data handling and concurrency, and a few critiqued the lack of descriptive comments in some examples. Overall, the archive was praised for its content, but there are areas that could be improved for greater user satisfaction.

The evaluation of the market-neutral pairs trading algorithm optimisation techniques focused on several aspects: speed, cache utilisation, statistical significance, and profitability. Speed analysis revealed that the optimised version of the trading algorithm significantly reduced latency by 87.38\%, reducing the mean time from 517,559ns to 65,588ns. Techniques like Inlining, SIMD, and Buffer Data Structure individually contributed to latency reduction by 21.41\%, 31.28\%, and 48.58\% respectively. The standard deviation in latency was much smaller in the optimised version (400 ns) compared to the non-optimised one (4233 ns), indicating more consistent performance. Cache analysis found that while all optimisation techniques increased the instruction count, their performances were influenced by cache utilisation. For instance, Buffer Data Structure had a cache miss rate of 19.089\% but significantly outperformed the original version with a 16.001\% cache miss rate. In contrast, Inlining had a similar cache miss rate of 19.151\% but still outperformed the original, highlighting the complexity of cache performance as a factor. Statistical significance was confirmed through a paired t-test, which showed large t-statistics and extremely low p-values, indicating that the latency improvements are not due to random variation but are statistically significant. Finally, the ultimate goal of these optimisations is to increase trading profit. Research indicates that HFT firms with faster trading capabilities tend to have higher profitability. This work also delves into the concept of Latency Arbitrage Opportunities (LAOs), which hinge on capturing price differentials between markets in small timeframes—something that optimised algorithms with reduced latency are better positioned to exploit. Given these multiple facets of analysis, there is compelling evidence that the optimisation techniques employed have significantly enhanced both the speed and effectiveness of the trading algorithm.

The performance evaluation of the Disruptor pattern compared to a traditional queue revealed significant advantages, particularly in speed and scalability. In speed tests, the Disruptor was nearly twice as fast as the Simple Queue, and its speed advantage increased as the number of events processed grew larger. Although the Disruptor initially showed a slightly higher cache-miss rate with small event sizes, it outperformed the Simple Queue in cache efficiency at larger scales. Statistical tests confirmed that the differences in latency were significant, with a t-statistic of 22.596 and an extremely low p-value. However, it's crucial to note that this evaluation mainly focused on speed and did not delve into other critical performance metrics like memory consumption and CPU load. Furthermore, while the benchmark employed a specific ``yield wait" strategy in the Disruptor, the impacts of different wait strategies on performance were not assessed, indicating areas for future comprehensive evaluation.

\subsection*{Future directions}

\subsubsection*{Repository expansion}
The repository holds the potential for expansive growth, encompassing a wider array of topics and techniques. The inclusion of advanced concepts like Variadic Templates could deepen the repository's coverage of modern C++ features. Moreover, delving into kernel bypass techniques could shed light on strategies for bypassing the operating system kernel to directly access network interfaces, potentially yielding substantial speed improvements in networking applications. Additionally, the repository could explore programming techniques related to networking and communication, offering insights into optimising data transmission, protocol handling, and synchronization mechanisms. By benchmarking these techniques, developers and researchers could gain a comprehensive understanding of their practical implications and select the most fitting solutions for their specific performance goals.

\subsubsection*{Live trading algorithm}
The pairs trading algorithm, though significantly optimised, could be taken a step further by coupling it with live market data feeds. Testing the algorithm in an actual trading environment could reveal the effectiveness of the optimisations under real-world conditions. Additionally if the OMS was connected to the trading logic then other programming strategies could be tested such as cache warming. Furthermore, expanding the repository to include a variety of trading algorithms—such as momentum trading, cross-exchange arbitrage, and trend-following strategies—would provide a comprehensive showcase of optimisation techniques across different trading scenarios. This expanded array of strategies could uncover unique challenges and opportunities for optimisation, enhancing the repository's practical value for algorithmic trading practitioners. 

\subsubsection*{Full trading system}
Exploring the integration of the Disruptor pattern with the pairs trading algorithm holds immense potential for benchmarking an entire trading system. By replacing the simple events in the Disruptor pattern with order objects, the combined system could simulate real trading scenarios. This integration would enable the benchmarking of both the Disruptor pattern's performance and the overall trading algorithm's speed improvement. This endeavor would be particularly intriguing as it merges high-performance computing principles with financial algorithms, providing a comprehensive view of how various optimisations synergize and impact the end-to-end trading process. The insights gained from this combined benchmarking could unveil novel approaches for accelerating algorithmic trading systems and offer guidance to developers seeking to create robust and efficient trading platforms.

In essence, the future directions for this work encompass a wide spectrum of possibilities. From expanding the repository's coverage to exploring real-world applications and combining optimisation techniques, each avenue promises to further our understanding of high-performance computing and its practical implications in diverse domains. Through continuous exploration, experimentation, and collaboration, this work has the potential to become an enduring resource for developers and researchers alike, shaping the landscape of optimisation and efficiency in technology and finance.

\bibliographystyle{abbrv}

\end{document}